\documentclass[aps,pra,twocolumn,10pt]{revtex4-2}
\usepackage{mathrsfs}
\usepackage{bbm}
\usepackage{amsmath}
\usepackage{amssymb}
\usepackage{amsthm}
\usepackage{graphicx}
\usepackage{mathtools}
\usepackage[italicdiff]{physics}
\usepackage{bbold}
\usepackage{bm}

\makeatletter
\usepackage{hyperref}
\usepackage{color}
\definecolor{supcol}{RGB}{10,50,180}
\definecolor{eqcol}{RGB}{220,10,100}
\hypersetup{
	colorlinks,
	citecolor=supcol,
	linkcolor=eqcol,
	urlcolor=supcol
}

\allowdisplaybreaks

\DeclareMathOperator{\Var}{Var}

\newcommand{\mca}{\mathcal}
\newcommand{\mbb}{\mathbb}
\newcommand{\mrm}{\mathrm}
\newcommand{\msf}{\mathsf}

\newcommand{\msc}{\mathscr}

\newcommand{\kvec}[1]{|#1)}
\newcommand{\bvec}[1]{(#1|}

\newcommand{\bkgvec}[2]{(#1|#2)}

\begin{document}
\title{Information-thermodynamic bounds on precision in interacting quantum systems}

\author{Ryotaro Honma}
\email{ryotaro.honma@yukawa.kyoto-u.ac.jp}
\affiliation{Center for Gravitational Physics and Quantum Information, Yukawa Institute for Theoretical Physics, Kyoto University, Kitashirakawa Oiwakecho, Sakyo-ku, Kyoto 606-8502, Japan}

\author{Tan Van Vu}
\email{tan.vu@yukawa.kyoto-u.ac.jp}
\affiliation{Center for Gravitational Physics and Quantum Information, Yukawa Institute for Theoretical Physics, Kyoto University, Kitashirakawa Oiwakecho, Sakyo-ku, Kyoto 606-8502, Japan}

\date{\today}

\begin{abstract}
The thermodynamic uncertainty relation quantifies a trade-off between the relative fluctuations of trajectory currents and the thermodynamic cost, indicating that the current precision is fundamentally constrained by entropy production. In classical bipartite systems, it has been shown that information flow between subsystems can enhance the current precision alongside thermodynamic dissipation. In this study, we investigate how information flow, local dissipation, and quantum effects jointly constrain current fluctuations within a subsystem of interacting quantum systems. Unlike classical bipartite systems, quantum subsystems can exhibit simultaneous state changes and maintain quantum coherence, which fundamentally alters the precision-dissipation trade-off. For this general setting, we derive a quantum thermokinetic uncertainty relation for interacting multipartite systems, establishing a thermodynamic trade-off between current fluctuations, information flow, local dissipation, and quantum effects. Our analysis shows that, in addition to local dissipation, both information exchange and quantum coherence play essential roles in suppressing current fluctuations. These results have important implications for the performance of quantum thermal machines, such as information-thermodynamic engines and quantum clocks. We validate our theoretical findings through numerical simulations on two representative models: an autonomous quantum Maxwell's demon and a quantum clock. These results extend uncertainty relations to multipartite open quantum systems and elucidate the functional role of information flow in fluctuation suppression.

\end{abstract}

\maketitle

\section{Introduction}
In mesoscale and nanoscale systems, physical states and observables inherently exhibit stochastic fluctuations. Among these, currents, such as electric, particle, heat, or work currents, play an essential role in characterizing the behavior of systems, particularly those driven far from equilibrium. These currents emerge as time-integrated observables associated with the flow of physical quantities and are key indicators of nonequilibrium processes. Recent advances in stochastic thermodynamics have uncovered fundamental constraints that govern the thermodynamic behavior of such fluctuating systems. In particular, trade-offs involving current fluctuations and entropy production, power and efficiency, operation speed and dissipation, as well as dynamical activity, have become active topics of study \cite{Horowitz.2020.NP,Pietzonka.2018.PRL,Shiraishi.2018.PRL,Maes.2020.PR,Vu.2023.PRX}.

\begin{figure}[b!]
\centering
\includegraphics[width=1\linewidth]{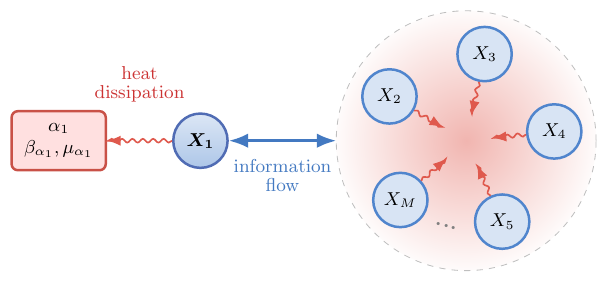}
\protect\caption{Schematic of interacting quantum systems. The system consists of $M$ subsystems $(X_1, \dots, X_M)$, where each subsystem $X_i$ is coupled to a set of reservoirs $\{\alpha_i\}$, characterized by inverse temperatures and chemical potentials $\{\beta_{\alpha_i}, \mu_{\alpha_i}\}$. Each subsystem dissipates heat into its environment and interacts with the others, giving rise to inter-subsystem information flow.}\label{fig:cover}
\end{figure}

A prominent result in this context is the thermodynamic uncertainty relation (TUR), which establishes a lower bound on the relative fluctuations of thermodynamic currents in terms of entropy production \cite{Barato.2015.PRL,Gingrich.2016.PRL,Dechant.2018.JSM,Hasegawa.2019.PRE,Hasegawa.2019.PRL,Timpanaro.2019.PRL,Koyuk.2020.PRL,Dieball.2023.PRL}. For classical Markov jump processes with even-parity variables, the TUR in steady states takes the form:
\begin{align}\label{eq.intro.TUR}
	\frac{\Var[\mca{J}]}{\ev{\mca{J}}^2} &\ge \frac{2k_B}{\Sigma},
\end{align}
where $\ev{\mca{J}}$ is the average of a time-integrated current $\mca{J}$, $\Var[\mca{J}]$ its variance, $\Sigma$ the total entropy production over a finite observation time, and $k_B$ the Boltzmann constant. This relation implies that achieving high precision in currents necessarily requires a large thermodynamic cost, manifested as entropy production.
A related but distinct bound is the kinetic uncertainty relation (KUR) \cite{Garrahan.2017.PRE,Terlizzi.2019.JPA,Prech.2025.PRX,Macieszczak.2024.arxiv}, which applies to arbitrary counting observables and relates their fluctuations to dynamical activity---a measure of the frequency of state transitions. The KUR reveals that, to suppress fluctuations of counting observables, the system must undergo transitions frequently, imposing a dynamical cost for precision. Furthermore, it has been shown that the TUR can be tightened by simultaneously considering both entropy production and dynamical activity, resulting in the thermokinetic uncertainty relation (TKUR) \cite{Vo.2022.JPA}. The TKUR provides a strictly stronger bound than both the TUR and KUR for thermodynamic currents, though it does not extend to arbitrary observables as the KUR does.

On the other hand, significant progress has also been made in understanding the role of information in physical systems \cite{Parrondo.2015.NP}. Information is intimately connected to physics in various ways, most notably through its relationship with stochastic thermodynamics via Shannon entropy. One of the most well-known examples where information plays a physically significant role is Maxwell's demon, a thought experiment that has inspired extensive theoretical and experimental investigations into the thermodynamics of information \cite{Maruyama.2009.RMP}.
A paradigmatic realization of Maxwell's demon is the Szilard engine \cite{Szilard.1929}, in which information exchange between a system of interest and a memory system through measurement and feedback can be used to reduce the system's entropy. More recent studies have extended these ideas to autonomous Maxwell's demons and autonomous bipartite systems, where information processing occurs internally, without the need for external measurement or control \cite{Horowitz.2014.NJP,Hartich.2014.JSM,Horowitz.2014.PRX,Horowitz.2015.JSM}.
In such bipartite systems, composed of two interacting subsystems, it has been shown that information flow naturally arises from the interactions \cite{Horowitz.2014.PRX}. This flow can significantly modify thermodynamic behavior: in particular, it can alter the second law and lead to situations where the entropy production of a subsystem becomes negative, compensated by information exchange with its counterpart. These developments collectively demonstrate that information flow is not merely an abstract concept but a key physical quantity that constrains and governs the thermodynamic limits of a system.

Building on this insight, recent studies have investigated the TUR for subsystems within interacting systems \cite{Vu.2020.PRR,Tanogami.2023.PRR}. In classical bipartite systems composed of two interacting subsystems, it has been shown that the presence of information flow between the subsystems can modify the original TUR bound. These results suggest that the fundamental limits on current fluctuations can be either relaxed or tightened, depending on the nature and extent of information exchange.
In parallel, it has been demonstrated that both the TUR and KUR can be violated in the quantum regime \cite{Agarwalla.2018.PRB,Ptaszynski.2018.PRB,Brandner.2018.PRL,Liu.2019.PRE,Saryal.2019.PRE,Cangemi.2020.PRB,Friedman.2020.PRB,Kalaee.2021.PRE,Menczel.2021.JPA,Bret.2021.PRE,Sacchi.2021.PRE,Lu.2022.PRB,Gerry.2022.PRB,Das.2023.PRE,Manzano.2023.PRR,Singh.2023.PRA,Farina.2024.PRE}, where uniquely quantum features, such as coherence, entanglement, and measurement backaction, fundamentally alter system dynamics. These effects undermine the assumptions underlying the classical bounds, necessitating new formulations. In response, several theoretical approaches have been developed to generalize these uncertainty relations to quantum systems \cite{Guarnieri.2019.PRR,Carollo.2019.PRL,Hasegawa.2020.PRL,Miller.2021.PRL.TUR,Vu.2022.PRL.TUR,Prech.2025.PRL,Vu.2025.PRXQ,Kwon.2025.CP,Yunoki.2025.arxiv,Nishiyama.2025.arxiv,Yoshimura.2025.arxiv,Hasegawa.2021.PRL,Ishida.2025.PRE,Vu.2025.arxiv}. Despite this progress, our understanding of thermodynamic precision constraints in the quantum regime remains incomplete, particularly in the presence of information flow \cite{QTRoadmap.2026.QST}. In this context, a natural and fundamental question remains open: How do information exchange and quantum effects jointly influence the precision-dissipation trade-offs in open quantum systems?
In this study, we address this question by deriving generalized uncertainty relations for open quantum systems composed of multiple interacting subsystems that exchange information. Our results reveal how inter-subsystem information flow and quantum effects fundamentally constrain the achievable precision of observables, thereby extending the thermodynamics of precision to a broader class of quantum systems.

Our central result is a quantum TKUR [Eq.~\eqref{eq:main.result.1}] for interacting multipartite systems composed of two or more subsystems $(X_1,X_2,\dots,X_M)$, governed by a master equation (see Fig.~\ref{fig:cover}). This result characterizes the precision of currents occurring within a single subsystem, revealing how information flow between the subsystem and the rest of the system, together with local dissipation and quantum effects, constrains the relative fluctuations of currents.
Notably, the derived bound holds for general non-steady (transient) states and is valid at arbitrary finite times, making it applicable to a broad class of nonequilibrium processes. As a corollary, we derive a quantum TUR [Eq.~\eqref{eq:main.result.2}] for interacting systems by applying additional inequality evaluations to the quantum TKUR. In contrast to the classical TUR for bipartite systems \cite{Tanogami.2023.PRR}, the quantum TUR includes both interaction-induced and genuinely quantum corrections in a unified correction term, reflecting the impact of nonclassical features on the current precision.
In addition, we explore the optimization of the quantum TKUR in the presence of multiple currents. By constructing an optimized current, namely, a linear combination of the given currents, we establish a multidimensional extension of the quantum TKUR [Eq.~\eqref{eq:main.result.3}], which provides the tightest possible bound.
As applications, we demonstrate that these results have significant implications for the performance of quantum thermal machines, such as information-thermodynamic engines and autonomous quantum clocks. Notably, they elucidate fundamental trade-offs between heat current fluctuations and efficiency, as well as between clock accuracy and dissipation.
To validate our theoretical findings, we perform numerical simulations on two representative models: an autonomous quantum Maxwell's demon \cite{Ptaszynski.2018.PRE,Ptaszynski.2019.PRL.QIF} and a quantum clock \cite{Erker.2017.PRX}. These examples highlight the theoretical relevance of our results and demonstrate how information flow and quantum coherence fundamentally reshape the thermodynamic limits on precision in open quantum systems.

\section{Setup}\label{sec:setup}
We consider an open quantum system composed of $M$ coupled subsystems $(X_1, X_2,\dots X_M)$, where each subsystem $X_i$ has finite Hilbert space dimension $d_i$ and is weakly coupled to a set of thermal reservoirs labeled by $\{\alpha_i\}$, each characterized by an inverse temperature $\beta_{\alpha_i}= 1/(k_BT_{\alpha_i})$ and a chemical potential $\mu_{\alpha_i}$.
Throughout this work, we set both the reduced Planck constant and the Boltzmann constant to unity for simplicity, i.e., $\hbar = k_B = 1$.
The total system Hamiltonian is given by
\begin{align}
	H_S = \sum_{i=1}^M H_{X_i} + H_{\rm int},
\end{align}
where $H_{X_i}$ denotes the Hamiltonian of subsystem $X_i$ for $i = 1, 2, \dots, M$, and $H_{\rm int}$ describes the interaction between subsystems.
Although our analysis can be readily extended to time-dependent dynamics, we restrict our focus to time-independent cases throughout this paper for notational simplicity.

Within Markovian approximations, we assume that the dynamics of the system is governed by the Gorini-Kossakowski-Sudarshan-Lindblad (GKSL) master equation \cite{Lindblad.1976.CMP,Gorini.1976.JMP}:
\begin{align}\label{eq:Lindblad.dyn}
	\dot{\varrho}_t&\coloneqq\mca{L}(\varrho_t)=-i\comm{H}{\varrho_t}+\sum_{i=1}^M\sum_{\alpha_i,k_i}\mca{D}[L_{\alpha_ik_i}]\varrho_t.
\end{align}
Here, $\varrho_t$ denotes the density matrix of the $M$ coupled subsystems at time $t$ and $\mca{D}[L]\circ\coloneqq(L\circ L^\dagger-\acomm{L^\dagger L}{\circ}/2)$ is the dissipator. Both the effective Hamiltonian $H$ and the jump operators $L_{\alpha_ik_i}$ are time-independent.
In principle, one can consider jump operators that act jointly on $X_i$ and other subsystems.
It is worth noting that the effective Hamiltonian $H$ may differ from the bare system Hamiltonian $H_S$ due to coupling with the environment. 
To formulate the thermodynamics of the system, we impose the local detailed balance condition \cite{Horowitz.2013.NJP,Manzano.2018.PRX}: each jump operator $L_{\alpha_ik_i}$ is associated with a corresponding reversed jump $L_{\alpha_ik_i^*}$, such that $L_{\alpha_ik_i}=e^{\Delta s_{\alpha_ik_i}/2}L_{\alpha_ik_i^*}^\dagger$. Here, $\Delta s_{\alpha_ik_i}$ denotes the entropy change in the environment induced by the jump $L_{\alpha_ik_i}$. For instance, if $L_{\alpha k}$ characterizes a transition between energy eigenstates, then the corresponding entropy change is given by $\Delta s_{\alpha k}=\beta_\alpha\Delta\epsilon$, where $\Delta\epsilon$ is the energy difference associated with the transition.

\subsection{Entropy production}
Here, we explain how entropy production is defined and introduce a decomposition relevant to interacting systems. In the framework of quantum thermodynamics \cite{Landi.2021.RMP}, the entropy production over a finite time interval $\tau$ is generally defined as the sum of the entropy changes in the system and its environment:
\begin{equation}
	\Delta S^{\rm tot}\coloneqq \Delta S+\Delta S^{\rm env},
\end{equation}
where $\Delta S\coloneqq -\tr(\varrho_\tau\ln\varrho_\tau) + \tr(\varrho_0\ln\varrho_0)$ and $\Delta S^{\rm env}\coloneqq \sum_{i=1}^M \sum_{\alpha_i} \beta_{\alpha_i} Q_{\alpha_i}$ represent, respectively, the total entropy change in the system and in the environment. Here, $Q_{\alpha_i}$ denotes the amount of heat dissipated into reservoir $\alpha_i$.
Notably, $\Delta S^{\rm env}$ can also be expressed in a more intuitive form as
\begin{equation}
	\Delta S^{\rm env} = \int_0^\tau \dd{t} \sum_{i=1}^M \sum_{\alpha_i, k_i} \tr(L_{\alpha_i k_i} \varrho_t L_{\alpha_i k_i}^\dagger) \Delta s_{\alpha_i k_i},
\end{equation}
which represents the cumulative entropy change in the environment due to quantum jumps.
Note that the definition of entropy production in quantum systems is not unique, particularly in contexts such as non-Abelian transport \cite{PRXQuantum.5.030355}. In this work, we adopt the conventional definition based on Ref.~\cite{Horowitz.2013.NJP}. Assuming the local detailed balance, this definition enjoys two desirable properties: (i) entropy production can be consistently decomposed into contributions from the system and its reservoirs, and (ii) a detailed fluctuation theorem for stochastic entropy production can be derived. From this perspective, it serves as a standard and physically meaningful definition for open quantum systems governed by the GKSL master equation.

We discuss a decomposition of entropy production. The entropy production rate can be written as $\dot S^{\rm tot} = \dot S + \dot S^{\rm env}$, where the individual contributions from the system and environment are given by
\begin{align}
	\dot S & = -\tr(\dot\varrho \ln \varrho), \label{eq:sys.ent} \\
	\dot S^{\rm env} & = \sum_{i, \alpha_i, k_i} \tr(L_{\alpha_i k_i} \varrho L_{\alpha_i k_i}^\dagger) \Delta s_{\alpha_i k_i}.
\end{align}
Here, the time dependence is omitted for notational simplicity.
Substituting $\dot\varrho = \mca{L}(\varrho)$ into Eq.~\eqref{eq:sys.ent}, we find that the entropy production rate admits the decomposition
\begin{equation}
	\dot S^{\rm tot} = \sum_{i=1}^M \dot S_i^{\rm tot},
\end{equation}
where each $\dot S_i^{\rm tot}$ corresponds to the partial entropy production rate associated with subsystem $X_i$, and is given by
\begin{align}
	\dot S_i^{\rm tot} \coloneqq \sum_{\alpha_i, k_i} \Big[&-\tr{(\mca{D}[L_{\alpha_i k_i}] \varrho_t) \ln \varrho_t}\notag\\
	&+\tr(L_{\alpha_i k_i} \varrho_t L_{\alpha_i k_i}^\dagger)\Delta s_{\alpha_i k_i}\Big].\label{eq:local.ent.prod}
\end{align}
Using the spectral decomposition $\varrho_t = \sum_n p_n(t) \dyad{n_t}$, we define $w^{mn}_{\alpha_i k_i}(t)\coloneqq|\mel{m_t}{L_{\alpha_i k_i}}{n_t}|^2$. Noting the relation $w_{\alpha_i k_i}^{mn} = e^{\Delta s_{\alpha_i k_i}} w_{\alpha_i k_i^*}^{nm}$ and performing algebraic manipulations, we obtain
\begin{align}
	\dot S_i^{\rm tot} & = \sum_{\alpha_i, k_i, m, n} w^{nm}_{\alpha_i k_i} p_m\ln \frac{w^{nm}_{\alpha_i k_i} p_m}{w^{mn}_{\alpha_i k_i^*} p_n} \notag \\
	& = \frac{1}{2} \sum_{\alpha_i, k_i, m, n} \qty(w^{nm}_{\alpha_i k_i} p_m - w^{mn}_{\alpha_i k_i^*} p_n) \ln \frac{w^{nm}_{\alpha_i k_i} p_m}{w^{mn}_{\alpha_i k_i^*} p_n}.
\end{align}
It is evident that $\dot S_i^{\rm tot} \ge 0$, which immediately implies the second law of thermodynamics, $\Delta S^{\rm tot} \ge 0$.

\subsection{Information flow}
Next, we describe the thermodynamics of quantum information flow between the coupled $M$ subsystems \cite{Ptaszynski.2019.PRL.QIF}. This framework can be viewed as a generalization of the classical theory developed for bipartite systems \cite{Horowitz.2014.PRX} to interacting quantum systems. 

A widely accepted definition of the quantum mutual information among interacting subsystems is given by \cite{5392532,PhysRevLett.104.080501}
\begin{equation}
	I_{1,\dots,M}\coloneqq \sum_{i=1}^{M} S_i - S,
\end{equation}
where $S_i$ denotes the von Neumann entropy of subsystem $X_i$, defined as $S_i \coloneqq -\tr(\varrho_{X_i} \ln \varrho_{X_i})$ with $\varrho_{X_i} = \tr_{\setminus i} \varrho_t$ being the reduced density matrix of subsystem $X_i$, and $S$ is the von Neumann entropy of the full composite system.
From an information-theoretic perspective, mutual information can be formulated using the relative entropy, defined as $D(\rho\|\sigma)\coloneqq\tr[\rho(\ln\rho-\ln\sigma)]$. In this case, the bipartite mutual information can be expressed as $I_{1,2}=D(\rho_t\|\rho_{X_1}\otimes\rho_{X_2})$, quantifying the total correlation between subsystems $X_1$ and $X_2$.
This definition naturally extends to the multipartite case, yielding $I_{1,\dots,M}=D(\rho_t\|\rho_{X_1}\otimes\cdots\otimes\rho_{X_M})$ for a system composed of $M$ subsystems \cite{5392532,PhysRevLett.104.080501}.
Taking the time derivative of $I_{1,\dots,M}$, we can decompose it into individual contributions from each subsystem:
\begin{equation}
	\dot I_{1,\dots,M} = \sum_{i=1}^M \dot I_i,
\end{equation}
where each $\dot I_i$ is given by
\begin{equation}
	\dot I_i \coloneqq \dot S_i + \sum_{\alpha_i,k_i} \tr{(\mca{D}[L_{\alpha_i k_i}] \varrho_t) \ln \varrho_t}.
\end{equation}
The quantity $\dot I_i$ represents the quantum information flow associated with subsystem $X_i$, quantifying how much information subsystem $X_i$ gains about the rest of the system.
From Eq.~\eqref{eq:local.ent.prod}, we note that
\begin{equation}
	\dot S_i^{\rm tot} = \sum_{\alpha_i} \beta_{\alpha_i} \dot Q_{\alpha_i} - \sum_{\alpha_i,k_i} \tr{(\mca{D}[L_{\alpha_i k_i}] \varrho_t) \ln \varrho_t},
\end{equation}
from which we immediately obtain the following equality:
\begin{align}
	\dot S_i^{\rm tot} &= \dot S_i + \sum_{\alpha_i} \beta_{\alpha_i} \dot Q_{\alpha_i} - \dot I_i.\label{eq:local.ent.prod.decom}
\end{align}
Here, the first two terms on the right-hand side represent the entropy changes in the subsystem and its local environment, and the last term accounts for information flow between the subsystem and the remainder of the system.
Defining the total entropy change associated with subsystem $X_i$ as $\dot \Sigma_i\coloneqq \dot S_i + \sum_{\alpha_i} \beta_{\alpha_i} \dot Q_{\alpha_i}$, Eq.~\eqref{eq:local.ent.prod.decom} can be rewritten as
\begin{equation}
    \dot S_i^{\rm tot}=\dot \Sigma_i - \dot I_i\ge 0,\label{eq:local.2nd.law}
\end{equation}
which can be interpreted as the second law of thermodynamics for subsystem $X_i$.
The information flow $\dot I_i$ plays a crucial role in this balance: it allows for the total entropy change of subsystem $X_i$ and its environment to become negative, provided the subsystem is exchanging information with the rest of the system.

\subsection{Quantum jump unraveling and currents}
The dynamics of a Markovian open quantum system can be unraveled into quantum jump trajectories. Defining the effective non-Hermitian Hamiltonian as $H_{\rm eff} \coloneqq H - (i/2) \sum_{i=1}^M \sum_{\alpha_i, k_i} L_{\alpha_i k_i}^\dagger L_{\alpha_i k_i}$, the GKSL master equation can be rewritten in the form:
\begin{equation}
	\dot\varrho_t = -i(H_{\rm eff} \varrho_t - \varrho_t H_{\rm eff}^\dagger) + \sum_{i=1}^M \sum_{\alpha_i, k_i} L_{\alpha_i k_i} \varrho_t L_{\alpha_i k_i}^\dagger.
\end{equation}
This representation shows that the GKSL dynamics can be interpreted as an ensemble of stochastic trajectories, where quantum jumps are induced by the operators $\{L_{\alpha_i k_i}\}$, while the no-jump evolution is governed by $H_{\rm eff}$.
In this picture, the system's density matrix at time $t$ is given by the average over conditioned density matrices, $\varrho_t = \mbb{E}[\psi_t]$, where the state $\psi_t$ evolves according to the stochastic Schr{\"o}dinger equation \cite{Breuer.2002}:
\begin{align}
	d\psi_t &= \mca{L}(\psi_t)dt + \sum_{i, \alpha_i, k_i} \qty( \frac{L_{\alpha_i k_i} \psi_t L_{\alpha_i k_i}^\dagger }{ \ev{L_{\alpha_i k_i}^\dagger L_{\alpha_i k_i}}_t  } - \psi_t )dN_{\alpha_i k_i, t} \notag\\
    &-\sum_{i, \alpha_i, k_i} \qty(L_{\alpha_i k_i} \psi_t L_{\alpha_i k_i}^\dagger - \ev{L_{\alpha_i k_i}^\dagger L_{\alpha_i k_i}}_t\psi_t )dt,
\end{align}
where $\ev{\circ}_t \coloneqq \tr(\circ\psi_t)$, and $dN_{\alpha_i k_i, t}$ is a stochastic increment taking the value $1$ if the jump $L_{\alpha_i k_i}$ occurs during $[t,t+dt]$, and $0$ otherwise. Its expectation satisfies $\mbb{E}[dN_{\alpha_i k_i, t}] = \ev{L_{\alpha_i k_i}^\dagger L_{\alpha_i k_i}}_t dt$.

We are particularly interested in currents occurring within a specific subsystem; for simplicity, we focus on $X_1$. For any stochastic trajectory determined by the set of jump events $\{dN_{\alpha_i k_i, t}\}$, we define a fluctuating current $\mca{J}_1$ as
\begin{equation}
	\mca{J}_1 \coloneqq \int_0^\tau \sum_{\alpha_1, k_1} c_{\alpha_1 k_1} dN_{\alpha_1 k_1, t},
\end{equation}
where $\{c_{\alpha_1 k_1}\}$ are real coefficients satisfying the time-antisymmetry condition $c_{\alpha_1 k_1} = -c_{\alpha_1 k_1^*}$.
Examples of such currents include the particle current, where $c_{\alpha_1 k_1} = 1$ for absorption events and $c_{\alpha_1 k_1} = -1$ for emissions, and the heat current, where $c_{\alpha_1 k_1} = \Delta q_{\alpha_1 k_1}$, with $\Delta q_{\alpha_1 k_1}$ denoting the heat exchanged with the environment during the jump $L_{\alpha_1 k_1}$.
The current average $\ev{\mca{J}_1}$ can be calculated as
\begin{align}
	\ev{\mca{J}_1}=\int_0^\tau \dd{t}\sum_{\alpha_1,k_1}c_{\alpha_1 k_1}\tr(L_{\alpha_1 k_1}\varrho_t L_{\alpha_1 k_1}^\dagger).
\end{align}
In general, higher-order moments of the current $\mca{J}_1$ can be computed using the method of full counting statistics \cite{Landi.2024.PRXQ}.
Defining the generating function $Z(u)\coloneqq \tr[e^{\mca{L}_u\tau}(\varrho_0)]$, the $n$th moment of $\mca{J}_1$ can be obtained via
\begin{align}
	\ev{\mca{J}_1^n}=\eval{(-i\partial_u)^n Z (u)}_{u=0}.
\end{align}
Here, the tilted superoperator $\mca{L}_u$ is given by
\begin{align}
	\mca{L}_u(\circ)&\coloneq \mca{L}(\circ)+\sum_{\alpha_1,k_1}(e^{iuc_{\alpha_1k_1}}-1)L_{\alpha_1k_1}\circ L_{\alpha_1k_1}^\dagger.
\end{align}

\section{Results}\label{sec:results}
In this section, we present our main results, the outline of the derivation, and their applications to quantum thermal machines.

\subsection{Quantum TKUR for interacting systems}
For interacting systems where subsystems exchange information and develop correlations, we focus on currents occurring within a {\it specific subsystem}, for instance, $X_1$. As our central result, we show that the relative fluctuation of such currents is constrained by the {\it partial} contributions of entropy production and dynamical activity associated with the subsystem. The relation is explicitly given by
\begin{align}
  \frac{\Var[\mca{J}_1]}{\ev{\mca{J}_1}^2}&\ge (1+\delta_{\mca{J}_1})^2\frac{4A_1}{(\Delta S_1^{\rm tot})^2}f\qty(\frac{\Delta S_1^{\rm tot}}{2A_1})^{2}.\label{eq:main.result.1}
\end{align}
Here, $f$ denotes the inverse function of $x\tanh(x)$, $\delta_{\mca{J}_1}$ is a correction term whose explicit form is given in Eq.~\eqref{eq:cur.avg.qcor}, and $\Delta S_1^{\rm tot}$ and $A_1$ are the partial entropy production and partial dynamical activity of subsystem $X_1$, defined as
\begin{align}
	\Delta S_1^{\rm tot}&\coloneqq\int_0^\tau\dd{t}\dot S_1^{\rm tot}(t),\\
	A_1&\coloneqq\int_0^\tau\dd{t}\sum_{\alpha_1,k_1}\tr(L_{\alpha_1k_1}\varrho_tL_{\alpha_1k_1}^\dagger).
\end{align}
The bound \eqref{eq:main.result.1} holds for arbitrary finite times and initial states. While we focus on quantum jump unraveling, an analogous result holds for quantum diffusion unraveling, as shown in Appendix \ref{app:qTKUR.qdu}.

It is worth noting that the quantum TKUR previously derived for system-wide currents in Ref.~\cite{Vu.2025.PRXQ} also applies to the local current $\mca{J}_1$ if one uses the {\it total} entropy production and dynamical activity. However, the bound developed here offers a more refined perspective: it demonstrates that the precision of local currents is constrained solely by {\it local} dissipation and activity.
As we elaborate below, this refinement is essential, as it uncovers the pivotal role of information flow between quantum subsystems in shaping the precision of local currents---a feature that remains hidden in previously established bounds.

To gain a clearer understanding of how information flow enhances current precision, we recall from Eq.~\eqref{eq:local.2nd.law} that the partial entropy production can be expressed as $\Delta S_1^{\rm tot} = \Sigma_1 - I_1$, where $I_1 \coloneqq \int_0^\tau \dd{t} \, \dot{I}_1(t)$ is the accumulated information flow between subsystem $X_1$ and the rest of the system.
By applying the inequality $f(x) \ge \sqrt{x}$, the quantum TKUR in Eq.~\eqref{eq:main.result.1} leads directly to a quantum TUR for subsystem $X_1$, given by
\begin{equation}\label{eq:main.result.2}
	\frac{\Var[\mca{J}_1]}{\ev{\mca{J}_1}^2} \ge \frac{2(1 + \delta_{\mca{J}_1})^2}{\Sigma_1 - I_1}.
\end{equation}
As shown, in addition to the total entropy change $\Sigma_1$ in subsystem $X_1$, the information flow $I_1$ appears explicitly in the denominator of the lower bound. This implies that the current precision can be enhanced, even when the local entropy production $\Sigma_1$ is small, provided that the information flow $-I_1$ is large.
In such cases, subsystem $X_1$ is effectively being monitored or ``learned about'' by the remainder of the system, namely, subsystems $(X_2, \dots, X_M)$ act as Maxwell's demon on $X_1$. This highlights the essential role of quantum information flow in enhancing precision.
It is also worth noting that, although the bound \eqref{eq:main.result.2} bears resemblance to the classical result derived in Ref.~\cite{Tanogami.2023.PRR}, the distinction lies in the correction term $\delta_{\mca{J}_1}$.
For quantum systems, $\delta_{\mca{J}_1}$ generally remains finite even in the fast-relaxation limit (see Appendix \ref{app:fast.relaxation} for details).
In our setting, interactions between $X_1$ and the rest of the system arise from both the interaction Hamiltonian $H_{\rm{int}}$ and the nonlocal nature of jump operators that act simultaneously on $X_1$ and other subsystems. When such interactions are absent, the corresponding contribution to $\delta_{\mca{J}_1}$ vanishes, leaving only quantum effects intrinsic to $X_1$. See Appendix \ref{app:special} for a detailed discussion of these limiting cases.
Therefore, the correction term $\delta_{\mca{J}_1}$ not only captures interaction-induced effects but also encodes the influence of quantum coherence.
In this sense, the inequality \eqref{eq:main.result.2} reveals how quantum information flow, combined with local dissipation and quantum coherence, fundamentally constrains the achievable precision of currents in interacting quantum systems.

It is instructive to examine how the present TKUR reduces to previously known results in several special cases. First, in the classical limit of a bipartite system ($M=2$), our bound \eqref{eq:main.result.2} reduces to the classical TUR derived in Ref.~\cite{Tanogami.2023.PRR}. Second, in the noninteracting case, where $H_{\mrm{int}}=\mbb{0}$ and the jump operators are purely local---of the form $L_{\alpha_1 k_1}=L_{\alpha_1k_1}'\otimes\mbb{1}_{\setminus 1}$---the dynamics of $X_1$ decouples from the rest. In this case, the new TKUR \eqref{eq:main.result.1} reduces to the single-system quantum TKUR previously established in Ref.~\cite{Vu.2025.PRXQ}. The special case $M=1$ also trivially corresponds to this single-system result. Detailed derivations of these reductions are provided in Appendix \ref{app:special}.

\subsection{Derivation of the central result \eqref{eq:main.result.1}}
Here, we provide an outline of the derivation of the main result \eqref{eq:main.result.1}; the detailed proof is deferred to Appendix \ref{app:proof.qtkur}. The proof strategy is based on the quantum Cram{\'e}r-Rao inequality.

We consider an auxiliary dynamics of the interacting system, parameterized by a virtual parameter $\theta$, such that the original dynamics \eqref{eq:Lindblad.dyn} is recovered in the limit $\theta = 0$. Specifically, the Hamiltonian and jump operators in the auxiliary dynamics are modified as
\begin{align}
	H_\theta &= H, \quad L_{\alpha_i k_i, \theta} = \sqrt{1 + \ell_{\alpha_i k_i}(t)\theta}L_{\alpha_i k_i},
\end{align}
where the coefficients $\{ \ell_{\alpha_i k_i}(t) \}$ are given by
\begin{align}
	\ell_{\alpha_1 k_1}(t) = \frac{ \tr(L_{\alpha_1 k_1} \varrho_t L_{\alpha_1 k_1}^\dagger) - \tr(L_{\alpha_1 k_1^*} \varrho_t L_{\alpha_1 k_1^*}^\dagger) }{ \tr(L_{\alpha_1 k_1} \varrho_t L_{\alpha_1 k_1}^\dagger) + \tr(L_{\alpha_1 k_1^*} \varrho_t L_{\alpha_1 k_1^*}^\dagger) },
\end{align}
and $\ell_{\alpha_ik_i}(t)=0$ for all $i>1$.
Here, the condition $|\theta| \ll 1$ must be satisfied to ensure the positivity of $1 + \ell_{\alpha_i k_i} \theta$. In other words, only the jump operators associated with subsystem $X_1$ are perturbed by $\theta$, while all others remain unchanged. Note that the jump operators in the auxiliary dynamics become time-dependent even when the original jump operators $\{L_{\alpha_i k_i}\}$ are time-independent, because they depend explicitly on the time-evolving state $\varrho_t$.

Within this auxiliary dynamics, we consider estimating the parameter $\theta$ from a time-integrated current $\mca{J}_1$, which may serve as a biased estimator. According to the quantum Cram{\'e}r-Rao inequality, the precision of this estimation is bounded from below by the quantum Fisher information:
\begin{align}
	\frac{ \Var[\mca{J}_1]_\theta}{ (\partial_{\theta}\ev{\mca{J}_1}_\theta)^2} \ge \frac{1}{\mca{I}_{\theta}}, \label{eq:Cramer-Rao1}
\end{align}
where the subscript $\theta$ denotes expectation values taken with respect to the auxiliary dynamics. Since the inequality \eqref{eq:Cramer-Rao1} holds for any $\theta$ satisfying $|\theta| \ll 1$, it is in particular valid at $\theta = 0$. Substituting $\theta = 0$ yields
\begin{align}
	\frac{ \Var[\mca{J}_1] }{ \qty( \eval{ \partial_{\theta} \ev{\mca{J}_1}_\theta }_{\theta = 0} )^2 } \ge \frac{1}{\mca{I}_0}. \label{eq:Cramer-Rao2}
\end{align}
The remaining task is to explicitly evaluate the quantum Fisher information $\mca{I}_0$ and the average term $\eval{ \partial_{\theta} \ev{\mca{J}_1}_\theta }_{\theta = 0}$.

For GKSL dynamics, the quantum Fisher information can be explicitly computed as
\begin{align}
	\mca{I}_0 = \int_0^\tau \dd{t} \sum_{\alpha_1, k_1} \ell_{\alpha_1 k_1}^2(t) \tr(L_{\alpha_1 k_1} \varrho_t L_{\alpha_1 k_1}^\dagger).
\end{align}
By applying the concavity of the function $(x^2/y)f(x/y)^{-2}$ and Jensen's inequality, we obtain an upper bound on the quantum Fisher information in terms of the partial entropy production and partial dynamical activity:
\begin{align}
	\mca{I}_0 &\le \frac{(\Delta S_1^{\rm tot})^2}{4 A_1} f\qty( \frac{\Delta S_1^{\rm tot}}{2 A_1} )^{-2}. \label{eq:qFI.ub}
\end{align}
The term $\eval{ \partial_\theta \ev{\mca{J}_1}_\theta }_{\theta = 0}$ can be calculated by expanding the density matrix $\varrho_{t,\theta}$ of the auxiliary dynamics as $\varrho_{t,\theta} = \varrho_t + \theta \varphi_t + O(\theta^2)$. Collecting the first-order terms yields a differential equation for the operator $\varphi_t$:
\begin{align}\label{eq:varphi}
	\dot{\varphi}_t = \mca{L}(\varphi_t) + \sum_{\alpha_1, k_1} \ell_{\alpha_1 k_1}(t) \mca{D}[L_{\alpha_1 k_1}] \varrho_t,
\end{align}
with the initial condition $\varphi_0 = \mbb{0}$.
Using this traceless operator $\varphi_t$, we can evaluate the derivative of the average current with respect to $\theta$ as
\begin{align}
	\eval{\partial_{\theta} \ev{\mca{J}_1}_\theta }_{\theta = 0} &= \ev{\mca{J}_1} + \ev{\mca{J}_1}_\varphi, \label{eq:par.avg.TUR}
\end{align}
where the second term is defined by
\begin{equation}
	\ev{\mca{J}_1}_\varphi \coloneqq \int_0^\tau \dd{t} \sum_{\alpha_1, k_1} c_{\alpha_1 k_1} \tr(L_{\alpha_1 k_1} \varphi_t L_{\alpha_1 k_1}^\dagger).
\end{equation}
Defining the correction term
\begin{equation}\label{eq:cur.avg.qcor}
\delta_{\mca{J}_1} \coloneqq \frac{\ev{\mca{J}_1}_\varphi}{\ev{\mca{J}_1}},
\end{equation}
we directly obtain the quantum TKUR \eqref{eq:main.result.1} for interacting quantum systems.
As shown, the correction term is governed by the operator $\varphi_t$, whose dynamics is driven by the total Hamiltonian and the set of jump operators. Interactions between subsystems arise in two ways: through the interaction Hamiltonian $H_{\rm int}$ and via nonlocal jump operators $\{L_{\alpha_ik_i}\}$ that act simultaneously on multiple subsystems. In the absence of both interactions and quantum coherence, this correction term vanishes for the steady state. Thus, $\delta_{\mca{J}_1}$ encapsulates the contributions from both inter-subsystem interactions and quantum coherence.
It is worth noting that $\delta_{\mca{J}_1}$ can remain nonzero for {\it nonstationary} initial states even in the classical, noninteracting limit due to relaxation dynamics.

\subsection{Multidimensional quantum TKUR}
Next, we consider the optimization of the quantum TKUR \eqref{eq:main.result.1} in the presence of multiple currents. It has been shown that incorporating multiple, potentially correlated, currents can lead to tighter uncertainty bounds than those derived from single-current TURs \cite{Dechant.2019.JPA,Vu.2019.PRE.UnderdampedTUR}. This improvement arises from the flexibility to construct an optimal current as a linear combination of the available currents, tailored to minimize the relative uncertainty. Such optimized bounds are particularly valuable in practical applications, such as the estimation of entropy production from experimental data. While one may restrict the analysis to mutually disjoint (non-overlapping) currents \cite{Moreira.2025.PRE}, we adopt a more general framework by allowing arbitrary linear combinations of multiple currents, without imposing any structural constraints.

Suppose we have a set of time-integrated currents $\{ \mca{J}_{1,1}, \dots, \mca{J}_{1,m} \}$, where, for example, each $\mca{J}_{1,i}$ is associated with a pair of forward and backward jumps $L_{\alpha_1 k_1}$ and $L_{\alpha_1 k_1^*}$ within subsystem $X_1$.
Based on this set, we define a new current $\mca{J}_1$ within subsystem $X_1$ as a linear combination of the given currents:
\begin{equation}
	\mca{J}_1\coloneqq\sum_{i=1}^Mz_i\mca{J}_{1,i},
\end{equation}
where $\vb*{z} \coloneqq [z_1, \dots, z_m]^\top$ is a vector of arbitrary real coefficients. 
Clearly, $\mca{J}_1$ is itself a valid time-integrated current and therefore satisfies the quantum TKUR bound \eqref{eq:main.result.1}:
\begin{align}\label{eq:tur.comb.curr}
  \frac{(\ev{\mca{J}_1}+\ev{\mca{J}_1}_\varphi)^2}{\Var[\mca{J}_1]} \le \frac{(\Delta S_1^{\rm tot})^2}{4A_1}f\qty(\frac{\Delta S_1^{\rm tot}}{2A_1})^{-2}.
\end{align}
A natural question arises: Which choice of coefficients $\{z_i\}$ yields the largest left-hand side in Eq.~\eqref{eq:tur.comb.curr}? 
This corresponds to optimizing the current $\mca{J}_1$ to achieve the tightest possible bound. 
Fortunately, this optimization problem can be analytically solved using the method of Lagrange multipliers \cite{Vu.2020.PRE}.
Defining the covariance matrix $\Xi$, where each element $\Xi_{ij}\coloneqq \ev{\mca{J}_{1,i} \mca{J}_{1,j}}-\ev{\mca{J}_{1,i}}\ev{\mca{J}_{1,j}}$ represents the correlation between currents $\mca{J}_{1,i}$ and $\mca{J}_{1,j}$, the variance of the current $\mca{J}_1$ can be written as $\Var[\mca{J}_1]=\vb*{z}^\top \Xi \vb*{z}$.
Since ${(\ev{\mca{J}_1}+\ev{\mca{J}_1}_\varphi)^2}/{\Var[\mca{J}_1]}$ is invariant under scalar rescaling of $\vb*{z}$, the optimization reduces to maximizing $\Var[\mca{J}_1]$ under the constraint $\ev{\mca{J}_1}+\ev{\mca{J}_1}_\varphi=1$. To proceed, define the Lagrangian 
\begin{align}
  \msc{L}(\lambda, \vb*{z})&=\Var[\mca{J}_1]-\lambda (\ev{\mca{J}_1}+\ev{\mca{J}_1}_\varphi-1)\notag\\
  &=\vb*{z}^\top \Xi \vb*{z}-\lambda (\vb*{z}^\top \vb*{\jmath}-1),
\end{align}
where the vector $\vb*{\jmath}$ is defined as
\begin{equation}
	\vb*{\jmath}\coloneqq\qty[\ev{\mca{J}_{1,1}}+\ev{\mca{J}_{1,1}}_\varphi,\dots,\ev{\mca{J}_{1,m}}+\ev{\mca{J}_{1,m}}_\varphi]^\top,
\end{equation}
and $\lambda$ is the Lagrange multiplier.
Solving $\partial\msc{L}/\partial z_i=0$ and $\partial \msc{L}/\partial \lambda =0$, we obtain the optimal solution:
\begin{align}
  \vb*{z}&=\frac{\Xi^{-1}\vb*{\jmath}}{\vb*{\jmath}^\top\Xi^{-1}\vb*{\jmath}},\\
  \lambda&=(\vb*{\jmath}^\top\Xi^{-1}\vb*{\jmath})^{-1}.
\end{align}
Substituting this result to Eq.~\eqref{eq:tur.comb.curr} yields the optimized quantum TKUR:
\begin{equation}\label{eq:main.result.3}
	\vb*{\jmath}^\top\Xi^{-1}\vb*{\jmath}\le \frac{(\Delta S_1^{\rm tot})^2}{4A_1}f\qty(\frac{\Delta S_1^{\rm tot}}{2A_1})^{-2}.
\end{equation}
This inequality is referred to as the multidimensional quantum TKUR for subsystems. Notably, it can also be independently derived from the multidimensional Cram{\'e}r-Rao inequality.
The bound is valid for any arbitrary set of currents $\{ \mca{J}_{1,i} \}$, and becomes tightest when the set forms a complete basis of currents. That is, each $\mca{J}_{1,i}$ counts the net jump associated with a specific pair of forward and backward operators $L_{\alpha_1 k_1}$ and $L_{\alpha_1 k_1^*}$, and the entire set collectively captures all quantum jump events occurring within subsystem $X_1$.

\subsection{Applications}
In what follows, we apply our results to two representative quantum thermal machines, highlighting how quantum effects and interactions between subsystems impose fundamental limits on their performance.

\subsubsection{Information-thermodynamic engines}
We consider an interacting quantum system in which each subsystem $X_i$ is coupled to a single reservoir $\alpha_i$, and the composite system operates as an information-thermodynamic engine, converting information into heat.
We focus on the case where subsystem $X_1$ is monitored or ``learned about'' by the remaining subsystems $\{X_i\}_{i \ge 2}$, such that the resulting information flow from $X_1$ to the others acts as a fuel source, driving a heat current from the reservoir $\alpha_1$ into $X_1$.
In this setting, we have $-\dot Q_{\alpha_1} \ge 0$ and $-\dot I_1 \ge 0$.
According to the generalized second law for subsystem $X_1$ [see Eq.~\eqref{eq:local.2nd.law}], the following inequality holds for steady states:
\begin{equation}
	- \dot I_1 \ge -\beta_{\alpha_1} \dot Q_{\alpha_1} \ge 0.
\end{equation}
This relation can be interpreted as the operational principle of information-thermodynamic engines: a portion of the quantum information flow is transformed into heat flow absorbed by subsystem $X_1$.
Motivated by this analogy, we define the efficiency of such an engine as
\begin{equation}
	\eta \coloneqq \frac{-\beta_{\alpha_1} \dot Q_{\alpha_1}}{-\dot I_1} \le 1,
\end{equation}
which quantifies the fraction of information flow converted into useful thermal energy.
Applying the quantum TUR \eqref{eq:main.result.2} to the rescaled heat current $\mca{J}_1 = -\beta_{\alpha_1} q_{\alpha_1}$, where $q_{\alpha_1}$ denotes the stochastic heat flow, we obtain the following trade-off relation among heat current, fluctuations, and efficiency:
\begin{equation}
	|\mca{J}_1| \le \frac{\Var[\mca{J}_1]}{2(1+\delta_{\mca{J}_1})^2} \frac{1 - \eta}{\eta}.
\end{equation}
In the classical case, and particularly in the fast-relaxation limit of subsystems $\{X_i\}_{i \ge 2}$, the correction term $\delta_{\mca{J}_1}$ vanishes \cite{Tanogami.2023.PRR}.
The fast-relaxation limit corresponds to a regime in which the subsystems $\{X_i\}_{i\ge 2}$ effectively act as an autonomous Maxwell's demon: their states quickly adjust in response to the state of $X_1$, thereby enabling a form of feedback control. In Appendix \ref{app:fast.relaxation}, we show that, in generic quantum systems, the correction term $\delta_{\mca{J}_1}$ can remain finite even in this limit.
This leads to a classical no-go statement: achieving maximal efficiency ($\eta \to 1$) at a finite heat current necessarily entails a divergence in current fluctuations.
In contrast, in the quantum regime, the presence of coherence can prevent $\delta_{\mca{J}_1}$ from vanishing in the fast-relaxation limit.
This opens the possibility of circumventing the classical constraint even in the fast-relaxation limit, namely, the divergence in fluctuations can be avoided if $|1 + \delta_{\mca{J}_1}|$ vanishes at the same order as $\sqrt{1 - \eta}$.
Hence, the derived trade-off relation reveals how quantum coherence fundamentally modifies classical thermodynamic bounds, offering a pathway toward designing high-performance information-thermodynamic engines that attain near-maximal efficiency at finite current, without incurring divergent fluctuations.
It is worth noting that similar performance may also be attainable in classical systems in the absence of a fast-relaxation limit.

\begin{figure*}[t]
\centering
\includegraphics[width=1\linewidth]{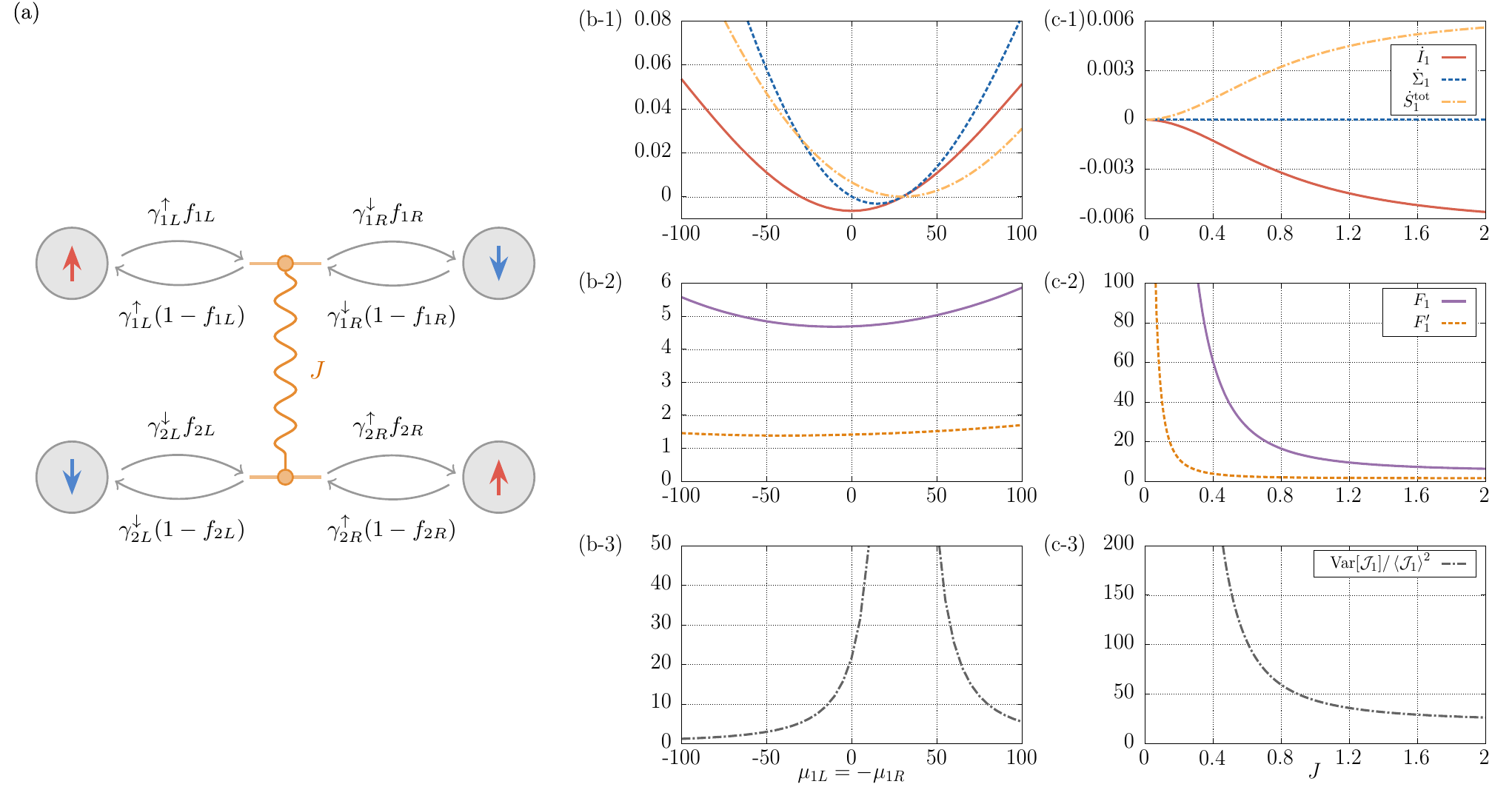}
\protect\caption{(a) Schematic illustration of the autonomous quantum Maxwell's demon, composed of two exchange-coupled quantum dots, each coupled to spin-polarized electronic leads arranged in an antiparallel configuration. (b), (c) Numerical verification of the quantum TUR in the steady state. In the upper panels, the solid, dashed, and dash-dotted lines represent the information flow rate $\dot I_1$, local entropy production rate $\dot \Sigma_1$, and partial entropy production rate $\dot S_1^{\rm tot}=\dot{\Sigma}_1 - \dot{I}_1 \ge 0$, respectively. In the middle panels, the solid and dashed lines correspond to the quality factors $F_1$ and $F_1'$, respectively. In the lower panels, dash-dotted lines depict the relative fluctuation $\Var[\mca{J}_1]/\ev{\mca{J}_1}^2$. The electrochemical potentials $\mu_{1L}=-\mu_{1R}$ or the exchange coupling $J$ is varied, whereas the remaining parameters are set as follows: $\beta = 0.01$, $J = 10$, $\mu_{1L} = -\mu_{1R}= 0$, $\mu_{2L} = -\mu_{2R} = -30$, $\gamma^{\downarrow}_{1L} = \gamma^{\uparrow}_{1R} = \gamma^{\uparrow}_{2L} = \gamma^{\downarrow}_{2R} = 0$, and $\tau = 10$. All remaining coupling strengths are set to $\gamma^{\sigma}_{i \nu} = 1$.}\label{fig:Maxwell}
\end{figure*}

\subsubsection{Autonomous quantum clocks}
Next, we consider an autonomous quantum clock, modeled as subsystem $X_1$, which interacts with the remaining subsystems $\{X_i\}_{i \ge 2}$ through information exchange. These auxiliary subsystems may act as a source of cyclic dynamics, driving the clock's ticking behavior. Among various measures of clock performance, a natural choice is the tick current $\mca{J}_1$, which counts the net number of ticks generated over time. The accuracy of the clock can then be quantified by the inverse Fano factor of this current \cite{Silva.2023.arxiv}:
\begin{equation}
\mca{N} \coloneqq \frac{\ev{\mca{J}_1}}{\Var[\mca{J}_1]}.
\end{equation}
This quantity remains finite in the long-time limit and increases with clock precision.
To characterize the thermodynamic cost of clock operation, we define the partial entropy production per tick as
\begin{equation}
	\sigma_{\rm tick} \coloneqq \frac{\Delta S_1^{\rm tot}}{\ev{\mca{J}_1}} = \frac{\Sigma_1 - I_1}{\ev{\mca{J}_1}}.
\end{equation}
While the local entropy production $\Sigma_1$ can become negative due to information flow exchanged between subsystems, the partial entropy production $\Delta S_1^{\mrm{tot}}$ is always nonnegative, as it incorporates both thermodynamic and informational contributions. Therefore, the latter provides a more appropriate measure of the thermodynamic cost in clock operation.
Applying the quantum TUR \eqref{eq:main.result.2} to the tick current immediately yields a trade-off between clock accuracy and thermodynamic cost:
\begin{equation}\label{eq:clock.pc}
\mca{N} \le \frac{\sigma_{\rm tick}}{2(1 + \delta_{\mca{J}_1})^2}.
\end{equation}
This inequality shows that clock accuracy is constrained not only by the partial entropy production $\sigma_{\rm tick}$ but also by the correction term $\delta_{\mca{J}_1}$, which captures the effects of quantum coherence and subsystem interactions. Notably, quantum clocks can significantly outperform classical counterparts if $|1 + \delta_{\mca{J}_1}| \to 0$, for instance, through the use of quantum coherent dynamics. In such cases, clock accuracy can scale exponentially with the entropy production, i.e., $\mca{N} \sim e^{\Omega(\sigma_{\rm tick})}$ \cite{Meier.2025.NP}, demonstrating a potential quantum supremacy in clock performance under thermodynamic constraints.
Conversely, if the clock precision scales as $\mathcal{N}=O(\sigma_{\mathrm{tick}}^\alpha)$ for some $\alpha>1$, then inequality \eqref{eq:clock.pc} implies $(1+\delta_{\mathcal{J}_1})^2 \le \sigma_{\mathrm{tick}}/O(\sigma_{\mathrm{tick}}^\alpha)\to 0$ in the large $\sigma_{\mathrm{tick}}$ limit. Therefore, achieving a superlinear scaling of $\mathcal{N}$ with respect to $\sigma_{\mathrm{tick}}$ requires that $|1+\delta_{\mathcal{J}_1}|\to 0$. Indeed, numerical simulations of a quantum many-body clock in Ref.~\cite{Meier.2025.NP} showed that $\mathcal{N}\sim n^{1.31}$ and $\sigma_{\mathrm{tick}}\sim \ln n$ can be achieved, where $n$ denotes the system size. As a result, this yields $(1+\delta_{\mathcal{J}_1})^2\le (\ln n)/ n^{1.31}\to 0$ as $n\to\infty$, indicating that $|1 + \delta_{\mathcal{J}_1}|$ asymptotically vanishes for large system sizes in this model. While a similar analysis can be performed using the TUR for the full system \cite{Vu.2025.PRXQ}, our focus is on how information flow, alongside local entropy production, constrains clock precision. This approach offers a more nuanced perspective by highlighting that precision is bounded by relevant local thermodynamic costs, rather than by the total cost of the full system, which may include contributions unrelated to the subsystem's performance.

\section{Numerical demonstration}\label{sec:num.demon}
In this section, we apply our results to two paradigmatic models, an autonomous quantum Maxwell's demon and a quantum clock, and numerically demonstrate the crucial role of quantum information flow in enhancing current precision.

\subsection{Autonomous quantum Maxwell's demon}
We first illustrate our results using an autonomous quantum Maxwell's demon model \cite{Ptaszynski.2018.PRE,Ptaszynski.2019.PRL.QIF}, which consists of two exchange-coupled single-level quantum dots. Each dot is weakly coupled to a pair of fully and collinearly spin-polarized leads arranged in an antiparallel configuration [see Fig.~\ref{fig:Maxwell}(a)].

The Hamiltonian of the isolated double-dot system (i.e., uncoupled from the leads) is given by
\begin{align}
  H&=H_0+H_{\rm{int}},\\
  H_0 &= \sum_{i \in \{1,2\}} \sum_{\sigma \in \{\uparrow, \downarrow\}} \epsilon_i d_{i\sigma}^\dagger d_{i\sigma} + \sum_{i \in \{1,2\}} U_i n_{i\uparrow} n_{i\downarrow} ,\\
  H_{\rm{int}}&= \frac{J}{2}(d_{1\uparrow}^\dagger d_{1\downarrow}d_{2\downarrow}^\dagger d_{2\uparrow}+d_{1\downarrow}^\dagger d_{1\uparrow}d_{2\uparrow}^\dagger d_{2\downarrow}),
\end{align}
where $\epsilon_i$ is the orbital energy of the $i$th quantum dot (i.e., subsystem $X_i$), $d_{i \sigma}^\dagger$ ($d_{i \sigma}$) is the creation (annihilation) operator for an electron with spin $\sigma$ in the $i$th dot, $n_{i \sigma}$ is the corresponding number operator, $U_i$ is the intradot Coulomb interaction energy in the $i$th dot, and $J$ denotes the exchange coupling between the dots.
The interaction Hamiltonian $H_{\rm{int}}$ induces spin exchange processes between subsystems $X_1$ and $X_2$. It can be equivalently expressed as $H_{\rm{int}} = J(S_1^x S_2^x + S_1^y S_2^y)$, where $S_i^\alpha$ represents the spin operator along the $\alpha$-direction in the $i$th dot. This form makes explicit that the model incorporates an XY-type exchange interaction, with no electron tunneling allowed between the dots.
Such interaction serves as the source of quantum information flow between the subsystems. Depending on the parameter regime, one subsystem may effectively function as a demon, monitoring and influencing the dynamics of the other.

We also assume that the intradot Coulomb interaction is sufficiently large to prevent double occupancy within each dot, which corresponds to the strong Coulomb blockade regime.
Under this condition, the state space of each subsystem can be spanned by the three basis states $\{\ket{0}, \ket{\uparrow}, \ket{\downarrow}\}$, where $\ket{0}$ denotes the empty dot, and $\ket{\uparrow}$ and $\ket{\downarrow}$ represent single-electron occupations with spin up and down, respectively.
The state space of the entire system is then given by the tensor product of the state spaces of the individual subsystems.

Each dot is connected to two leads, labeled as $i\nu$, where $i \in\{1,2\}$ specifies the dot to which the electrode is coupled, and $\nu = L~(R)$ indicates the left (right) lead.
The electrochemical potential and temperature of lead $i\nu$ are denoted by $\mu_{i\nu}$ and $T_{i\nu}$, respectively, with the corresponding inverse temperature given by $\beta_{i\nu} = 1/T_{i\nu}$.
Assuming weak coupling and well-separated energy levels, the transport dynamics can be described by the GKSL master equation \cite{Ptaszynski.2018.PRE,Ptaszynski.2019.PRL.QIF}:
\begin{align}
  \dot{\varrho}_{t}=-i[H, \varrho_t]&+\sum_{i,\nu, \sigma} \gamma_{i \nu}^\sigma f_{i \nu} \mca{D}[d_{i\sigma}^\dagger]\varrho_t\notag\\
  &+\sum_{i,\nu, \sigma} \gamma_{i \nu}^{\sigma} (1-f_{i \nu}) \mca{D}[d_{i\sigma}]\varrho_t,
\end{align}
where $\varrho_t$ is the density operator of the total system, $\gamma_{i \nu}^{\sigma}$ are constants characterizing the transition rates, and $f_{i \nu} = 1/[e^{\beta_{i\nu}(\epsilon_i-\mu_{i\nu})}+1]$ denotes the Fermi-Dirac distribution for electrons in lead $i\nu$.
Note that the dissipative terms satisfy the local detailed balance condition: $\gamma_{i \nu}^{\sigma} f_{i \nu}/[\gamma_{i \nu}^{\sigma} (1-f_{i \nu})] = \exp[-\beta_{i\nu}(\epsilon_i - \mu_{i\nu})]$.

In the following analysis, we assume that all reservoirs are at the same temperature, i.e., $T_{i\nu} = T$ for all $i,\nu$, and set $\gamma^{\downarrow}_{1L} = \gamma^{\uparrow}_{1R} = \gamma^{\uparrow}_{2L} = \gamma^{\downarrow}_{2R} = 0$.
We focus on a particle current $\mca{J}_1$ within subsystem $X_1$ that assigns $+1$ to each jump associated with $d^\dagger_{1 \uparrow}$, $-1$ to each jump associated with $d_{1 \uparrow}$, and $0$ to all other jumps. This current is proportional to the net heat current flowing from reservoir $1L$ into subsystem $X_1$.
To verify the validity of the main result \eqref{eq:main.result.2} and to benchmark it against existing results, we introduce the following quality factors:
\begin{align}
	F_{1}&\coloneqq\frac{\Var[\mca{J}_1]}{\ev{\mca{J}_1}^2(1+\delta_{\mca{J}_1})^2}(\Sigma_1-I_1)\ge 2,\label{eq:F1.def}\\
	F_{1}'&\coloneqq\frac{\Var[\mca{J}_1]}{\ev{\mca{J}_1}^2}(\Sigma_1-I_1).\label{eq:F1p.def}
\end{align}
By definition, the condition $F_{1}\ge 2$ confirms the validity of the bound \eqref{eq:main.result.2}, while a value $F_{1}' < 2$ signifies a violation of the conventional TUR applied to subsystem $X_1$.

\begin{figure}[t]
\centering
\includegraphics[width=1\linewidth]{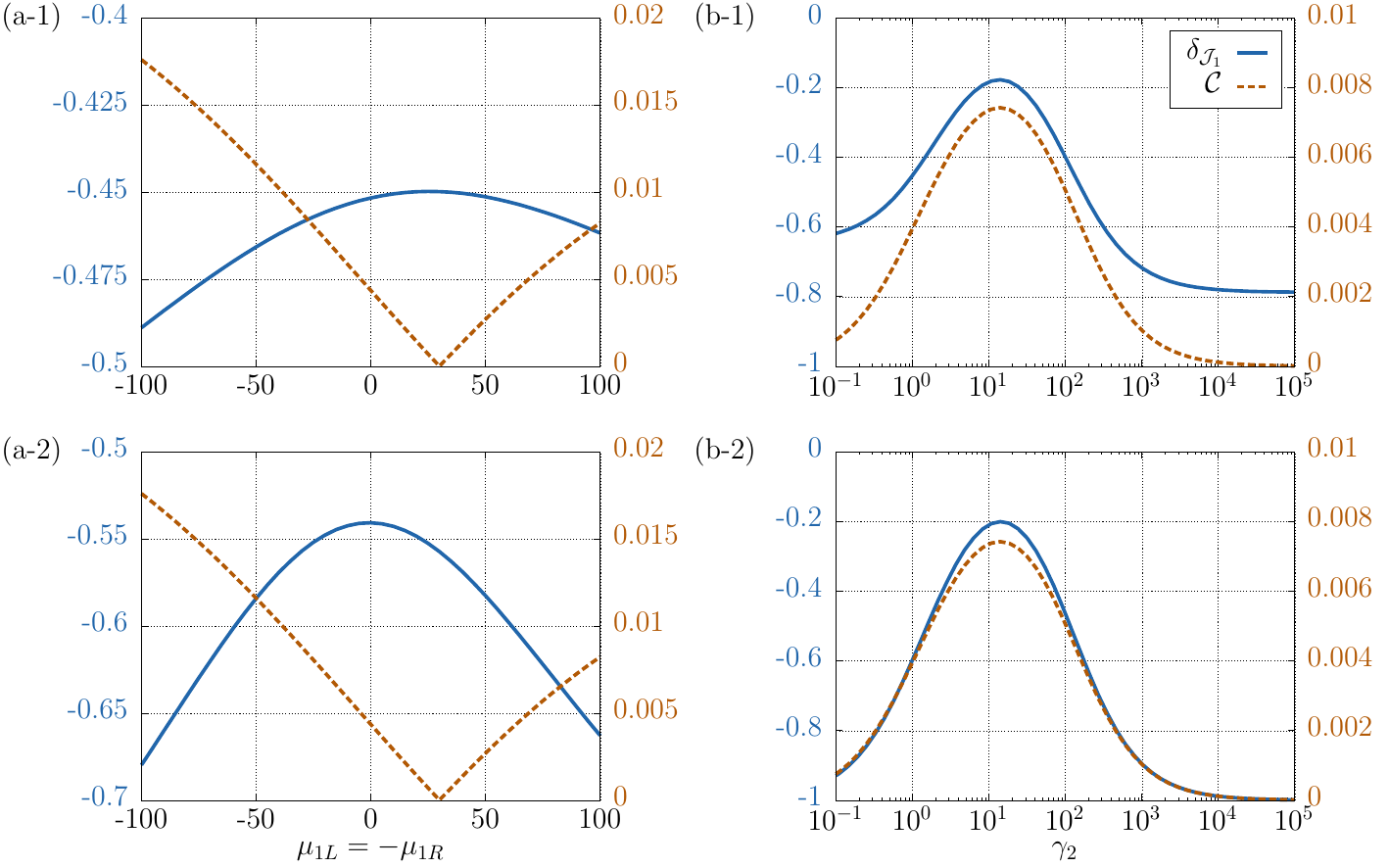}
\protect\caption{Numerical plots of the $l_1$-norm of quantum coherence $\mca{C}$ (dashed lines) and the correction term $\delta_{\mca{J}_1}$ (solid lines): (a) As functions of the electrochemical potentials $\mu_{1L} = -\mu_{1R}$ with fixed period $\tau=10$ in (a-1) and the long-time limit $\tau\rightarrow \infty$ in (a-2); (b) As functions of the coupling strength $\gamma_2$ with $\tau=10$ in (b-1) and $\tau\rightarrow \infty$ in (b-2). All other parameters are fixed as in Fig.~\ref{fig:Maxwell}.}\label{fig:Maxwell.coh}
\end{figure}

\begin{figure*}[t]
\centering
\includegraphics[width=1.0\linewidth]{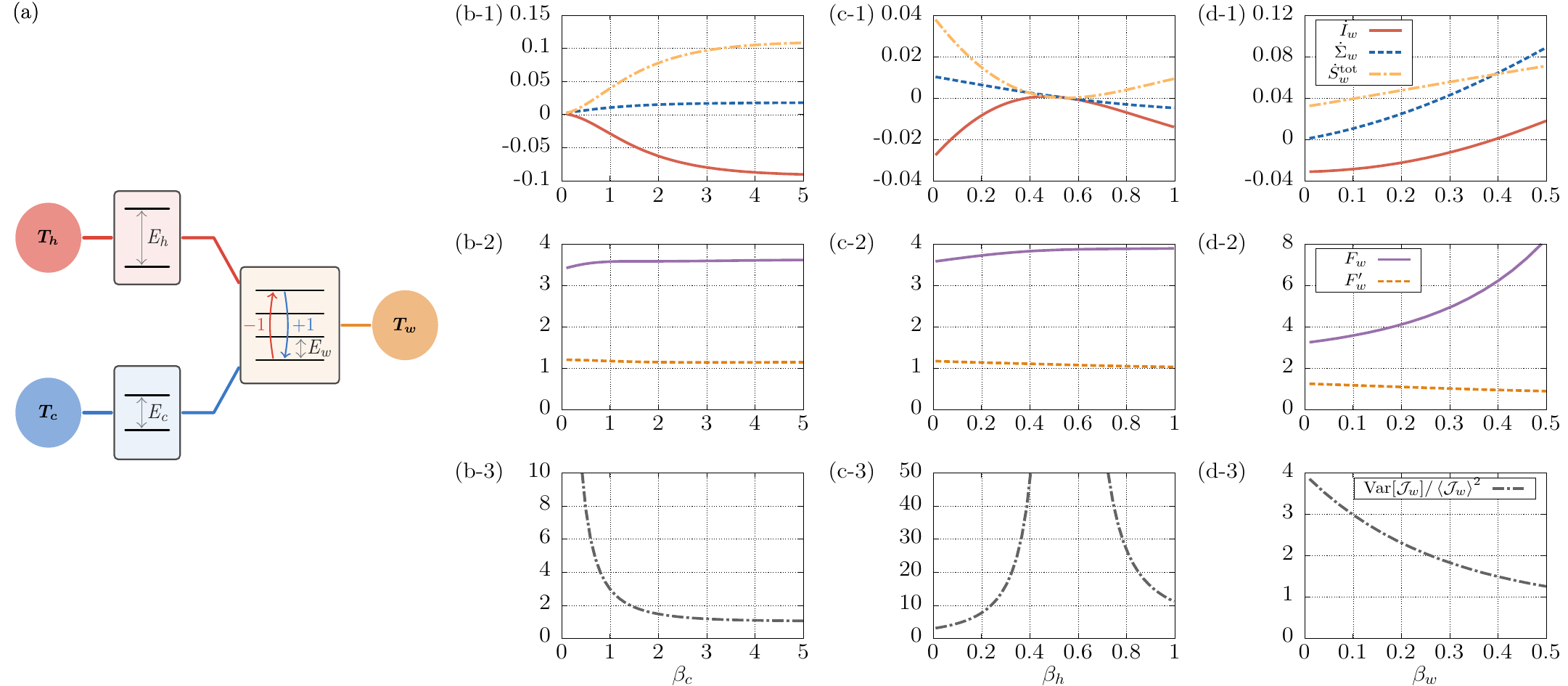}
\protect\caption{(a) Schematic of the autonomous quantum clock, composed of a two-qubit heat engine and a four-level ladder, each coupled to thermal reservoirs. (b), (c), (d) Numerical validation of the quantum TUR \eqref{eq:main.result.2} in the steady state. In the upper panels, the solid, dashed, and dash-dotted lines correspond to the information flow rate $\dot I_w$, local entropy production rate $\dot \Sigma_w$, and partial entropy production rate $\dot S_w^{\rm tot}=\dot\Sigma_w-\dot I_w\ge 0$, respectively. 
In the middle panels, the solid and dashed lines represent the quality factors $F_w$ and $F_w'$, respectively. In the lower panels, the dash-dotted lines correspond to the relative fluctuation $\Var[\mca{J}_w]/\ev{\mca{J}_w}^2$.
The inverse temperatures are varied as indicated, while the remaining parameters are fixed: $\beta_c=1$, $\beta_h=10^{-3}$, $\beta_w=0.1$, $E_c=E_w=1$, $E_h=E_c+E_w$, $g=5$, $\gamma_c=\gamma_h=1$, $\gamma_w=0.1$, and $\tau=10$.}\label{fig:clock}
\end{figure*}

We consider the system in its stationary state, where a steady information flow between subsystems $X_1$ and $X_2$ occurs.
Motivated by experimental considerations, we analyze the system's behavior over a finite-time interval $\tau$, rather than taking the asymptotic long-time limit.
The electrochemical potential $\mu_{1L}~(=-\mu_{1R})$ is varied, while all other parameters are held fixed.
The fluctuation of the current $\mca{J}_1$ is numerically evaluated using the method of full counting statistics.
The local dissipation $\dot\Sigma_1$, the information flow $\dot I_1$, the partial entropy production rate $\dot S_{1}^{\rm tot}$, the relative fluctuation $\Var[\mca{J}_1]/\ev{\mca{J}_1}^2$, and the quality factors $F_{1}$ and $F_{1}'$ are computed and plotted in Figs.~\ref{fig:Maxwell}(b-1) and \ref{fig:Maxwell}(b-2).
At the specific parameter setting $\mu_{1L} = 30~(= -\mu_{2L})$, we observe that all quantities $\dot{I}_1$, $\dot{\Sigma}_1$, and $\dot S_1^{\rm tot}$ vanish, and the relative fluctuation diverges.
This indicates that the system reaches thermal equilibrium and no net current flows in this case.
In general, the thermal Gibbs state $\pi\propto\exp[-\beta (H_0-\sum_{i,\nu,\sigma} \mu_{i\nu} n_{i\sigma})]$ becomes the stationary equilibrium state when the condition $\mu_{1L}-\mu_{1R}=-(\mu_{2L}-\mu_{2R})$ is fulfilled, where $\sum_{i,\nu,\sigma} \mu_{i\nu} n_{i\sigma}=\mu_{1L}n_{1\uparrow}+\mu_{1R}n_{1\downarrow}+\mu_{2L}n_{2\downarrow}+\mu_{2R}n_{2\uparrow}$.
This fact can be derived analytically by using the local detailed balance conditions together with the commutation relation $[H_{\mrm{int}}, \sum_{i,\nu,\sigma} \mu_{i\nu} n_{i\sigma}]=(J/2)\{(-\mu_{1L}+\mu_{1R}-\mu_{2L}+\mu_{2R})(d_{1\uparrow}^\dagger d_{1\downarrow}d_{2\downarrow}^\dagger d_{2\uparrow}-\text{h.c.})\}$.
Importantly, we identify parameter regimes where the information flow $\dot{I}_1$ becomes negative, indicating that subsystem $X_1$ is being monitored by subsystem $X_2$, which thereby acts as a quantum Maxwell's demon.
Throughout the full range of $\mu_{1L}$, $F_{1}\ge 2$ is satisfied, confirming the validity of the bound \eqref{eq:main.result.2}.
On the other hand, $F_{1}'$ is smaller than $2$, indicating the violation of the conventional TUR and the necessitate of the correction term $\delta_{\mca{J}_1}$.

To further elucidate the essential role of information flow, we now fix all system parameters except the exchange coupling strength $J$, which we vary. The resulting behavior of the relevant thermodynamic quantities is presented in Figs.~\ref{fig:Maxwell}(c-1) and \ref{fig:Maxwell}(c-2).
As shown in Fig.~\ref{fig:Maxwell}(c-1), the local entropy production rate $\dot{\Sigma}_1$ remains nearly zero across all values of $J$, indicating negligible local dissipation. 
This behavior arises from the absence of local thermodynamic forces: with $\mu_{1L}=\mu_{1R}=0$, no net heat current flows into subsystem $X_1$, leading to vanishing local entropy production.
Nonetheless, a nonzero information flow from subsystem $X_1$ to $X_2$ persists throughout, revealing a regime in which the fluctuation of the current is constrained purely by information exchange via $H_{\mathrm{int}}$, independent of entropy production.
Moreover, Fig.~\ref{fig:Maxwell}(c-2) shows that as the coupling strength $J$ increases, the magnitude of the information flow grows correspondingly. This enhanced flow leads to a marked improvement in current precision, as evidenced by the reduced relative fluctuation.
These numerical results clearly illustrate the pivotal role of quantum information flow in determining the thermodynamic performance of interacting quantum systems. Specifically, they demonstrate that information flow alone can regulate current fluctuations, even in the absence of local entropy production, thereby highlighting its foundational role in precision control within quantum thermodynamics.

Finally, we investigate the role of quantum coherence in shaping the precision of currents. Previous studies have shown that the violation of the TUR cannot be fully explained by the quantum coherence present in the quantum states alone \cite{Kalaee.2021.PRE}, but rather by the quantum coherent dynamics of the system \cite{Vu.2022.PRL.TUR}. In the context of interacting quantum systems, the relevant contribution from coherent dynamics is captured by the correction term $\delta_{\mca{J}_1}$.
To examine this effect more clearly, we quantify the coherence in the steady state using the $l_1$-norm of coherence \cite{Baumgratz.2014.PRL}, defined as
\begin{equation}
	\mca{C}\coloneqq\sum_{m\neq n}|\mel{m}{\varrho^{\rm ss}}{n}|^2.
\end{equation}
Here, $\varrho^{\rm ss}$ denotes the steady state and the basis states $\ket{m}$ take the form $\ket{\msf{s}_1\msf{s}_2}$, where each $\msf{s}_i\in\{\uparrow,\downarrow,0\}$.
As illustrated in Fig.~\ref{fig:Maxwell.coh}(a), the coherence $\mca{C}$ vanishes at the equilibrium point $\mu_{1L}=30$ and increases monotonically as the system deviates from equilibrium, signaling the buildup of quantum coherence in the steady state. This increase in coherence is strongly correlated with the decrease of the correction term $\delta_{\mca{J}_1}$, which remains within the range $[-1,0]$ in our data, though in principle it can lie outside this interval. Consequently, this reduction suppresses $(1 + \delta_{\mca{J}_1})^2$ in the TUR bound \eqref{eq:main.result.2}.
In Fig.~\ref{fig:Maxwell.coh}(b), we further analyze the effect by varying the coupling strength of the second quantum dot, considering both the finite-time regime [Fig.~\ref{fig:Maxwell.coh}(b-1)] and the long-time limit [Fig.~\ref{fig:Maxwell.coh}(b-2)]. As $\gamma_2$ increases, the second dot approaches the fast-relaxation limit. In the long-time limit, the correction term can be analytically calculated using the Moore-Penrose pseudo-inverse (see Appendix \ref{app:fast.relaxation} for details). While in the classical case $\delta_{\mca{J}_1}$ is known to vanish in these limits, we observe that it remains strictly negative even at large $\gamma_2$, indicating that the quantum coherent dynamics continues to contribute nontrivially to the correction term.

\subsection{Quantum clock}
Next, we exemplify our results using an autonomous quantum clock consisting of a two-qubit heat engine and a ladder subsystem driven by the engine dynamics \cite{Erker.2017.PRX}. Each of the two engine qubits and the ladder is coupled to its own thermal reservoir [see Fig.~\ref{fig:clock}(a)].
Specifically, the two heat-engine qubits, with energy gaps $E_h$ and $E_c$, are connected to hot and cold reservoirs at temperatures $T_h$ and $T_c$, respectively. The ladder, a $d$-dimensional system with equally spaced energy levels of spacing $E_w$, is coupled to a third reservoir at temperature $T_w$.
The ladder serves as the clockwork, producing ticks via spontaneous decays from its highest-energy level to its ground state. Let $\ket{1}_i$ ($\ket{0}_i$) denote the excited (ground) state of qubit $i \in\{h,c\}$, and let $\ket{k}_w$ denote the $k$-th level of the ladder. The free Hamiltonian of the composite system (two qubits and ladder) is given by
\begin{align}
	H_0 &= \sum_{i=h,c} E_i n_i +\sum_{k =0}^{d-1}k E_w n_k,
\end{align}
where $n_i = \ket{1}_i \bra{1}_i$ for $i \in\{h, c\}$, and $n_k = \ket{k}_w \bra{k}_w$ for $k = 0, 1, \dots, d-1$.
The interaction Hamiltonian governing the coupling between the engine and the ladder is
\begin{align}
	H_{\rm int} = g \sum_{k=1}^{d-1} (\sigma_c^\dagger \sigma_h \sigma_k^\dagger + \sigma_c \sigma_h^\dagger \sigma_k),
\end{align}
where the qubit lowering operators are defined as $\sigma_i = \ket{0}_i \bra{1}_i$ for $i\in\{h, c\}$, and the ladder lowering operator is $\sigma_k = \ket{k-1}_w \bra{k}_w$ for $k = 1, 2, \dots, d-1$.
This interaction allows the engine to increment or decrement the ladder's energy level, enabling transitions between adjacent levels $k$ and $k+1$. However, it does not facilitate direct transitions between the highest and lowest energy levels of the ladder. Instead, the periodic reset of the ladder from its top level to the ground state, which generates the clock ticks, is mediated exclusively by its coupling to the thermal reservoir, as detailed below.
The total Hamiltonian of the system is defined as
\begin{equation}
	H = H_0 + H_{\rm int},
\end{equation}
and we impose the energy conservation constraint $E_h = E_c + E_w$, ensuring that energy exchanged among the subsystems is conserved.

Assuming weak coupling and invoking the Born and Markov approximations, the dissipative dynamics induced by the thermal reservoirs can be described by the GKSL master equation. The density operator $\varrho_t$ of the total system evolves according to
\begin{align}
  \dot \varrho_t=-i[H, \varrho_t]
  &+\qty(\gamma_w \mca{D}[\sigma_w]+\gamma_w e^{\beta_w (d-1)E_w} \mca{D}[\sigma_w^\dagger])\varrho_t\notag\\
  &+\qty(\gamma_h \mca{D}[\sigma_h]+\gamma_h e^{-\beta_h E_h} \mca{D}[\sigma_h^\dagger])\varrho_t \notag\\
  &+\qty(\gamma_c \mca{D}[\sigma_c]+\gamma_c e^{-\beta_c E_c} \mca{D}[\sigma_c^\dagger])\varrho_t,
\end{align}
where $\sigma_w=\ket{d-1}_w\bra{0}_w$ denotes the ladder's decay operator, and $\gamma_h$, $\gamma_c$, $\gamma_w$ are the dissipative coupling strengths associated with the hot, cold, and ladder reservoirs, respectively.
Note that only transitions between the highest and lowest energy levels of the ladder are mediated by the ladder reservoir. Such selective coupling can be implemented experimentally, for instance, in optical systems using a frequency filter placed between the system and the reservoir \cite{Kalaee.2021.PRE,PhysRevLett.2.262}. The filter permits interaction only with photons in a narrow frequency range corresponding to a specific energy gap, effectively suppressing all other transitions. In the present quantum clock model, where the reservoir couples exclusively to the extremal ladder levels, this form of reservoir engineering offers a feasible physical realization.
Each dissipative term satisfies the local detailed balance condition, thereby ensuring the thermodynamic consistency.

In the following analysis, we set the ladder dimension to $d = 4$ for numerical simulations. We focus on the tick current $\mca{J}_w$, which counts $+1$ for each jump induced by $\sigma_w^\dagger$ (representing a transition from the highest to the lowest energy level of the ladder), $-1$ for the reverse jump $\sigma_w$, and $0$ for all other transitions. This current effectively captures the net number of ticks generated by the clock mechanism.
To verify the validity of the main results, we investigate the system in the steady state and explore how the current statistics change when varying the inverse temperatures $\beta_c$, $\beta_h$, or $\beta_w$, while keeping all other parameters fixed. The relative fluctuation of the tick current $\Var[\mca{J}_w]/\ev{\mca{J}_w}^2$ is numerically computed using the method of full counting statistics.
Likewise, we restrict our analysis to the finite-time regime, which is more pertinent for experimental implementations.
All numerical results are presented in Fig.~\ref{fig:clock}.

As shown in Fig.~\ref{fig:clock}(b-1), the magnitude of the information flow increases monotonically as the cold reservoir temperature decreases (i.e., as $\beta_c$ increases), while the local dissipation remains small and negligible in comparison. This identifies a regime where information flow dominates over entropy production, emphasizing its central role in the system's nonequilibrium behavior.
Simultaneously, Fig.~\ref{fig:clock}(b-2) shows that the relative fluctuation of the tick current $\mca{J}_w$ decreases monotonically with increasing $\beta_c$, indicating that information flow effectively suppresses fluctuations even in the absence of significant local dissipation.
We define the quality factors $F_w$ and $F_w'$ for the current $\mca{J}_w$ analogously to Eqs.~\eqref{eq:F1.def} and \eqref{eq:F1p.def}.
Within the entire examined range of $\beta_c$, the quality factor $F_w$ consistently satisfies $F_w \ge 2$, thereby numerically confirming the validity of the quantum TUR given in Eq.~\eqref{eq:main.result.2}. In contrast, the conventional quality factor $F_w'$, which neglects the effects of quantum coherence and subsystem interactions, remains strictly below $2$, signaling a violation of the original TUR.
These results demonstrate the crucial role of quantum effects and inter-subsystem interactions in precision enhancement. 

Figures \ref{fig:clock}(c-1) and \ref{fig:clock}(c-2) reveal the presence of a specific parameter setting in which all relevant quantities, namely, the information flow $\dot{I}_w$, the local entropy production rate $\dot{\Sigma}_w$, the correction term $\delta_{\mca{J}_w}$, simultaneously vanish. As a result, the relative fluctuation of the tick current diverges, indicating that the system is in equilibrium.
In general, the system's equilibrium state has the Gibbs form $\pi=e^{-H_\beta}/\tr e^{-H_\beta}$ when the thermodynamic balance condition $\beta_c E_c-\beta_h E_h+ \beta_w E_w=0$ is satisfied. Here, $H_\beta\coloneqq \sum_{i=h,c} \beta_i E_i n_{i}+\beta_w\sum_{k=0}^{d-1} k E_w n_{k}$.
This fact can be verified analytically by using the local detailed balance conditions and the commutation relation $[H_{\mrm{int}}, H_\beta]=-g(\beta_c E_c-\beta_h E_h+ \beta_w E_w)\sum_{k=1}^{d-1}(\sigma_c^\dagger \sigma_h \sigma_k^\dagger-\text{h.c.})$.
Throughout the entire range of $\beta_h$ considered, the quality factor $F_w$ remains above $2$, verifying the validity of the quantum TUR \eqref{eq:main.result.2}, while $F_w'<2$, again signaling the breakdown of the conventional TUR.

\begin{figure}[t]
\centering
\includegraphics[width=1\linewidth]{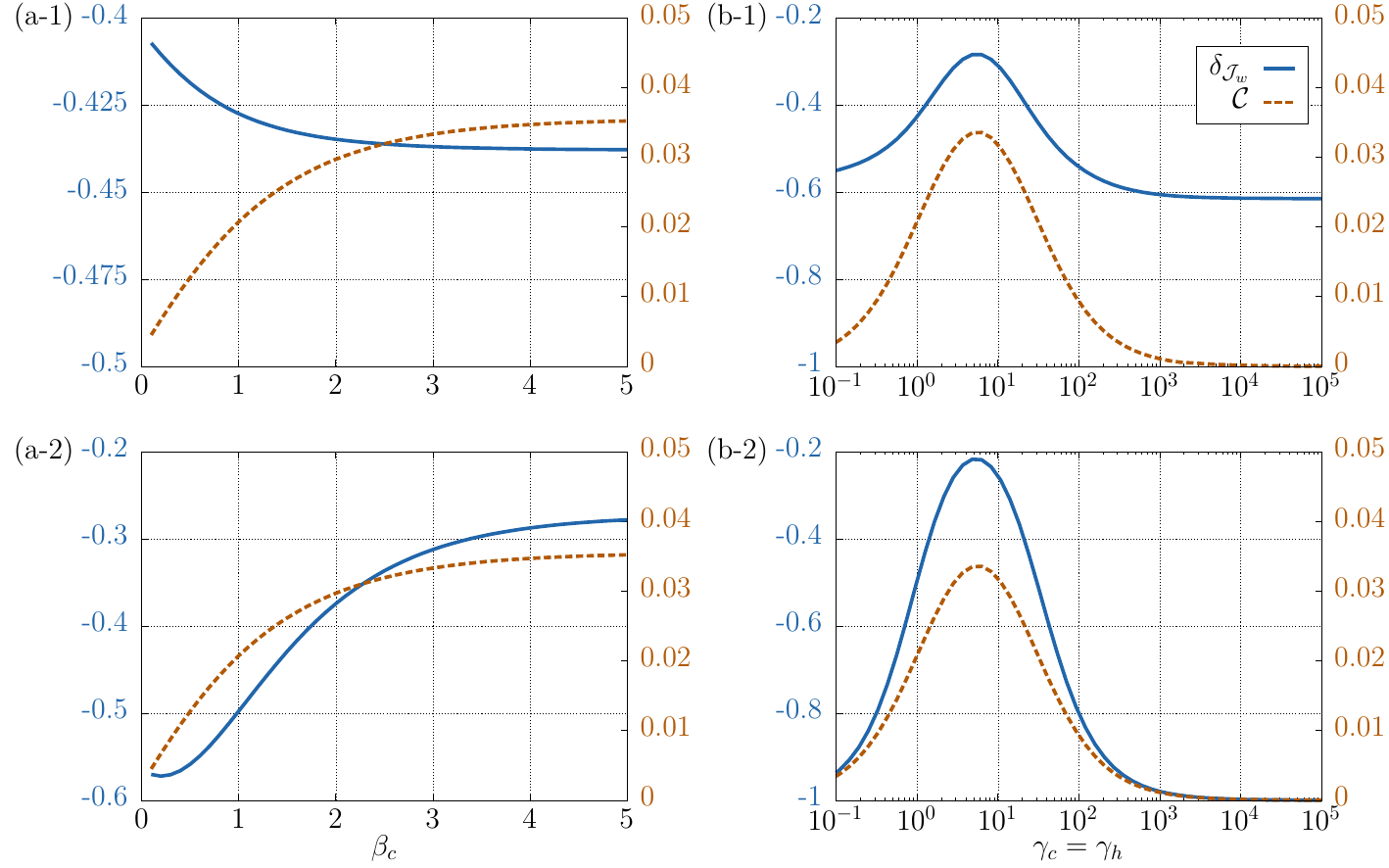}
\protect\caption{Numerical plots of the $l_1$-norm of quantum coherence $\mca{C}$ (dashed lines) and the correction term $\delta_{\mca{J}_w}$ (solid lines): (a) As functions of the inverse temperature $\beta_c$ with $\tau=10$ in (a-1) and $\tau\rightarrow \infty$ in (a-2); (b) As functions of the coupling strengths $\{\gamma_c, \gamma_h\}$ with $\tau=10$ in (b-1) and $\tau\rightarrow \infty$ in (b-2). All other parameters are fixed as in Fig.~\ref{fig:clock}.}\label{fig:clock.coh}
\end{figure}

To further investigate, we vary $\beta_w$ and observe that the precision of the tick current improves as the temperature $T_w$ decreases. This observation aligns with the fact that lower temperatures of the clockwork reservoir suppress the reverse transitions from the ground state to the highest energy level of the ladder, thereby enhancing the regularity of tick generation and, consequently, clock accuracy. In the low-temperature regime, the local entropy production per tick becomes significantly larger and surpasses the contribution of information flow, thus playing the dominant role in constraining current fluctuations.

As a final analysis, we examine the effect of quantum coherence on current precision. Analogously to the previous example of Maxwell's demon, we quantify coherence using the $l_1$-norm in the basis $\{\ket{i_c i_h i_w}\}$, where $i_c, i_h \in \{0,1\}$ and $i_w \in \{0, \dots, d-1\}$.
We vary the inverse temperature $\beta_c$ and the coupling strengths $\gamma_c = \gamma_h$, and plot both the coherence $\mca{C}$ and the correction term $\delta_{\mca{J}_w}$ in Fig.~\ref{fig:clock.coh}. Once again, a strong correlation is observed between the coherence and the correction term, as shown in Fig.~\ref{fig:clock.coh}(a), where $\mca{C}$ increases while $\delta_{\mca{J}_w}$ decreases.
In Fig.~\ref{fig:clock.coh}(b), as both $\gamma_c$ and $\gamma_h$ increase, the engine dynamics approaches the fast-relaxation limit. Nevertheless, the correction term remains nonvanishing, indicating the persistent contribution of quantum coherent dynamics.

\section{Conclusion}\label{sec:con.dis}
In this study, we derived information-thermodynamic bounds on the relative fluctuations of currents in interacting quantum systems, where subsystems continuously exchange information. Our central result is the quantum TKUR \eqref{eq:main.result.1}, which establishes a lower bound on the fluctuations of time-integrated currents within a subsystem, expressed in terms of local entropy production, local dynamical activity, and information flow between the subsystem and the rest of the system. This bound is broadly applicable to quantum dynamics governed by GKSL master equations that satisfy the local detailed balance condition, and it holds at arbitrary finite times, making it suitable for both steady and transient nonequilibrium processes.

As a direct consequence, we derived the quantum TUR \eqref{eq:main.result.2}, which reveals how entropy production and information flow jointly constrain current precision in the presence of quantum coherence. Remarkably, our result implies that information flow alone can suppress current fluctuations, even in the absence of entropy production, thereby highlighting the essential role of information flow in nonequilibrium quantum thermodynamics. This generalizes the classical result obtained for bipartite systems in Ref.~\cite{Tanogami.2023.PRR} to multipartite quantum systems, where subsystems may evolve simultaneously and interact coherently. In contrast to the classical bound, the correction term in the quantum TUR encapsulates not only interaction-induced effects but also genuine quantum coherence, thereby demonstrating how quantum effects influence precision limits in open quantum systems.

We further extended our analysis to derive a multidimensional quantum TKUR \eqref{eq:main.result.3}, incorporating multiple currents and their correlations to yield the tightest possible bound. This multidimensional framework provides a unified approach to optimizing current precision in interacting quantum systems.

Finally, we illustrated the significance of our results through two representative applications: information-thermodynamic engines and autonomous quantum clocks. In both cases, we demonstrated how the fundamental performance limits of these quantum thermal machines are shaped by the interplay of dissipation, coherence, and information flow.

As a promising direction for future work, it will be important to explore the role of information flow in quantum systems under measurement and feedback control, where additional layers of information processing and back-action may further refine or reshape the thermodynamic limits established here.

\begin{acknowledgments}
The authors are grateful to Hisao Hayakawa and Keiji Saito for fruitful discussions.
This work was supported by JSPS KAKENHI Grant No.~JP23K13032.
\end{acknowledgments}

\section*{Data availability}
The data are not publicly available upon publication. The data are available from the authors upon reasonable request.

\appendix

\section{Detailed derivation of the central result \eqref{eq:main.result.1}}\label{app:derivation}
Here we provide a complete derivation of our central result [Eq.~\eqref{eq:main.result.1}], which proceeds in three steps:
\begin{enumerate}
	\item[(1)] Generalized quantum Cram{\'e}r-Rao inequality: In Appendix \ref{app:gen.CR.ine}, we derive the generalized quantum Cram{\'e}r-Rao inequality \eqref{eq:Cramer-Rao1} for GKSL dynamics:
\begin{equation}\label{eq:gen.CR.ine}
	\frac{\Var[\mca{J}_1]_\theta}{(\partial_\theta\ev{\mca{J}_1}_\theta)^2}\ge\frac{1}{\mca{I}_{\theta}}.
\end{equation}
	
	\item[(2)] Calculation of quantum Fisher information: In Appendix \ref{app:quantum.Fisher}, we show how the quantum Fisher information $\mca{I}_{\theta}$ can be explicitly computed for GKSL dynamics.
	
	\item[(3)] Evaluation under perturbation: In Appendix \ref{app:proof.qtkur}, we evaluate $\mca{I}_0$ and $\partial_\theta\ev{\mca{J}_1}_\theta|_{\theta=0}$ for the perturbed dynamics.
\end{enumerate}
Finally, in Appendix \ref{app:qTKUR.qdu}, we demonstrate that an analogous result can be obtained for the case of quantum diffusion unraveling.
Hereafter, the notation of the subsystem's current $\mca{J}_1$ is simplified to $\mca{J}$.

\subsection{Generalized quantum Cram{\'e}r-Rao inequality}\label{app:gen.CR.ine}
For convenience, we consider a general GKSL dynamics parameterized by a scalar parameter $\theta$,
\begin{equation}\label{eq:gen.mas}
	\dot\varrho_{t,\theta}=-i[H_{\theta},\varrho_{t,\theta}]+\sum_{k\ge 1}\mca{D}[L_{k,\theta}]\varrho_{t,\theta},
\end{equation}
where the initial state is independent of $\theta$, i.e., $\varrho_{0,\theta}=\varrho_0$.
For simplicity, we consider an initial pure state $\varrho_0=\dyad{\psi_0}$, as any mixed initial state can be purified using an ancillary system, and the same analysis can be applied straightforwardly.
We discretize the total evolution time as $\tau=N \Delta t$, where $N$ is a positive integer, and the continuous-time limit can be achieved by taking the limit $N \rightarrow \infty$.
In the short-time limit, the master equation \eqref{eq:gen.mas} can be expressed in the form of the Kraus representation as
\begin{equation}\label{eq:Kraus.rep}
	\varrho_{t+\Delta t,\theta}=\sum_{k\ge 0}M_{k,\theta}\varrho_{t,\theta}M_{k,\theta}^\dagger,
\end{equation}
where the Kraus operators $\{M_{k,\theta}\}$ satisfy the completeness condition $\sum_{k\ge 0}M_{k,\theta}^\dagger M_{k,\theta}=\mbb{1}_S$.
It is worth noting that the unraveling is not unique.
In the case of quantum jump unraveling, the Kraus operators are given by
\begin{equation}
	M_{0,\theta}=\mbb{1}_S-iH_{\rm eff,\theta}\Delta t,~M_{k,\theta}=L_{k,\theta}\sqrt{\Delta t},
\end{equation}
where the effective Hamiltonian is defined as $H_{\rm eff,\theta}\coloneqq H_\theta-(i/2)\sum_{k\ge 1}L_{k,\theta}^\dagger L_{k,\theta}$.
The Kraus representation in Eq.~\eqref{eq:Kraus.rep} enables us to unravel the deterministic GKSL dynamics into an ensemble of stochastic trajectories $\{\Gamma = (k_1, \dots, k_N)\}$, where each operator $M_{k_i,\theta}$ accounts for the evolution during the time interval $[(i-1)\Delta t, i\Delta t]$.
The jump events and the final pure state of the system for all trajectories can be encoded into the following extended state:
\begin{equation}
	\ket{\Psi_{\tau,\theta}}\coloneqq\sum_{\Gamma} M_{k_{N},\theta}\dots M_{k_1,\theta}\ket{\psi_0}\otimes \ket{\Gamma},
\end{equation}
where the virtual ancilla states $\ket{\Gamma}\coloneqq\ket{k_{N},\dots,k_1}$ record jump information and form an orthonormal basis.
For each trajectory $\Gamma$, the current $\mca{J}$ can be calculated as
\begin{equation}
	\mca{J}=\mca{J}(\Gamma)\coloneqq\sum_{i=1}^{N}c_{k_i},
\end{equation}
where we set $c_0=0$.
We then define the self-adjoint current operator:
\begin{equation}
	\msf{J}\coloneqq\mbb{1}_S\otimes\sum_{\Gamma}\mca{J}(\Gamma)\dyad{\Gamma},
\end{equation}
so that the average and variance of the current can be expressed as
\begin{align}
	\ev{\mca{J}}_\theta&=\mel{\Psi_{\tau,\theta}}{\msf{J}}{\Psi_{\tau,\theta}},\\
	\Var[\mca{J}]_\theta&=\mel{\Psi_{\tau,\theta}}{(\msf{J}-\ev{\mca{J}}_\theta)^2}{\Psi_{\tau,\theta}}.
\end{align}
We define the pure state density operator $\Psi_{\tau,\theta}\coloneqq\dyad{\Psi_{\tau,\theta}}$ and introduce the Symmetric Logarithmic Derivative (SLD) operator $\msf{L}_\theta$ the via
\begin{equation}
	\partial_\theta{\Psi_{\tau,\theta}}=\frac{1}{2}\qty( \msf{L}_\theta{\Psi_{\tau,\theta}}+{\Psi_{\tau,\theta}}\msf{L}_\theta ).
\end{equation}
The quantum Fisher information is then defined using the SLD operator as
\begin{equation}\label{eq:qFI.def}
	\mca{I}_{\theta}\coloneqq\tr(\msf{L}_\theta^2{\Psi_{\tau,\theta}})=\mel{\Psi_{\tau,\theta}}{\msf{L}_\theta^2}{\Psi_{\tau,\theta}}.
\end{equation}
Applying the Cauchy-Schwarz inequality, we obtain
\begin{align}
	|\partial_\theta\ev{\mca{J}}_\theta|&=|\tr(\msf{J}\partial_\theta{\Psi_{\tau,\theta}})|\notag\\
	&=|\tr{(\msf{J}-\ev{\mca{J}}_\theta)\partial_\theta {\Psi_{\tau,\theta}}}|\notag\\
	&=\frac{1}{2}|\mel{\Psi_{\tau,\theta}}{(\msf{J}-\ev{\mca{J}}_\theta)\msf{L}_\theta+\msf{L}_\theta(\msf{J}-\ev{\mca{J}}_\theta)}{\Psi_{\tau,\theta}}|\notag\\
	&\le \sqrt{\mca{I}_{\theta}\Var[\mca{J}]_\theta},
\end{align}
which leads directly to the generalized Cram{\'e}r-Rao inequality:
\begin{equation}\label{eq:app.Cramer-Rao1}
	\frac{\Var[\mca{J}]_\theta}{(\partial_\theta\ev{\mca{J}}_\theta)^2}\ge\frac{1}{\mca{I}_{\theta}}.
\end{equation}
It is worth emphasizing that this inequality holds for arbitrary trajectory observables, extending beyond simple counting-type observables.

\subsection{Calculation of quantum Fisher information}\label{app:quantum.Fisher}
Next, we describe the calculation of quantum Fisher information $\mca{I}_{\theta}$ for GKSL dynamics, following the approach in Ref.~\cite{Gammelmark.2014.PRL}.
Taking the derivative of pure state $\Psi_{\tau,\theta}$ with respect to parameter $\theta$ yields $\partial_\theta\Psi_{\tau,\theta} =\dyad{\partial_\theta \Psi_{\tau,\theta}}{\Psi_{\tau,\theta}}+\dyad{\Psi_{\tau,\theta}}{\partial_\theta \Psi_{\tau,\theta}}$.
From this relation and the normalization condition $\braket{\Psi_{\tau,\theta}}{\Psi_{\tau,\theta}}=1$, we can verify that the SLD operator can be expressed as
\begin{align}
  \msf{L}_\theta= 2(\dyad{\partial_\theta \Psi_{\tau,\theta}}{\Psi_{\tau,\theta}}+\dyad{\Psi_{\tau,\theta}}{\partial_\theta \Psi_{\tau,\theta}}).
\end{align}
Substituting this expression into Eq.~\eqref{eq:qFI.def}, we obtain
\begin{align}
  \mca{I}_{\theta}&=4\qty( \braket{\partial_\theta \Psi_{\tau,\theta}}{\partial_\theta \Psi_{\tau,\theta}} - |\braket{\Psi_{\tau,\theta}}{\partial_\theta \Psi_{\tau,\theta}}|^2 ).
\end{align}
Here, we use the fact that $\braket{\partial_\theta \Psi_{\tau,\theta}}{\Psi_{\tau,\theta}}$ is a pure imaginary number, which can be derived from the identity $0=\partial_\theta\braket{\Psi_{\tau,\theta}}=\braket{\partial_\theta \Psi_{\tau,\theta}}{\Psi_{\tau,\theta}}+\braket{\Psi_{\tau,\theta}}{\partial_\theta \Psi_{\tau,\theta}}$.
Notably, the quantum Fisher information can be equivalently written as
\begin{align}
  \mca{I}_{\theta}&=4\partial_{\theta_1\theta_2}^2(\ln|\braket{\Psi_{\tau,\theta_1}}{\Psi_{\tau,\theta_2}}|)|_{\theta_1=\theta_2=\theta}.\label{eq:qFI.equiv}
\end{align}
To verify this, recall that for any complex function $\phi(x)$,
\begin{align}
  \partial_x\ln |\phi(x)|=\Re\qty[\frac{\partial_x \phi(x)}{\phi(x)}].
\end{align}
Then,
\begin{align}
  &4\eval{\partial_{\theta_1\theta_2}^2 (\ln|\braket{\Psi_{\tau,\theta_1}}{\Psi_{\tau,\theta_2}}|)}_{\theta_1=\theta_2=\theta}\notag\\
  &=4\eval{\partial_{\theta_1} \Re 
  \qty(\frac{\braket{\Psi_{\tau,\theta_1}}{\partial_{\theta_2} \Psi_{\tau,\theta_2}}}{\braket{\Psi_{\tau,\theta_1}}{\Psi_{\tau,\theta_2}}})}_{\theta_1=\theta_2=\theta}\notag\\
  &=4\qty(\braket{\partial_{\theta}\Psi_{\tau,\theta}}{\partial_{\theta} \Psi_{\tau,\theta}}-|\braket{\Psi_{\tau,\theta}}{\partial_\theta \Psi_{\tau,\theta}}|^2)\notag\\
  &=\mca{I}_{\theta}.
\end{align}
Now, note that $|\braket{\Psi_{\tau,\theta_1}}{\Psi_{\tau,\theta_2}}|=|\tr(\dyad{\Psi_{\tau,\theta_1}}{\Psi_{\tau,\theta_2}})|=|\tr(\varrho_{\tau,\vb*{\theta}})|$, where the two-sided operator $\varrho_{\tau,\vb*{\theta}}$ is defined as
\begin{equation}
	\varrho_{\tau,\vb*{\theta}}\coloneqq\sum_{\Gamma} \msf{M}_{\Gamma,\theta_1}\varrho_0\msf{M}_{\Gamma,\theta_2}^\dagger,
\end{equation}
where $\msf{M}_{\Gamma,\theta}\coloneqq M_{k_{N},\theta}\dots M_{k_1,\theta}$.
By definition, the time evolution of the operator $\varrho_{t,\vb*{\theta}}$ can be described as
\begin{equation}\label{eq:two.side.op.evol}
	\varrho_{t+\Delta t,\vb*{\theta}}=\sum_{k\ge 0}M_{k,\theta_1}\varrho_{t,\vb*{\theta}}M_{k,\theta_2}^\dagger.
\end{equation}
Taking the limit $\Delta t\to 0$ of Eq.~\eqref{eq:two.side.op.evol} yields the differential equation that describes the time evolution of the two-sided operator $\varrho_{t,\vb*{\theta}}$,
\begin{align}\label{eq:two.side.master.eq}
	\dot\varrho_{t,\vb*{\theta}}&=-i(H_{\theta_1}\varrho_{t,\vb*{\theta}}-\varrho_{t,\vb*{\theta}}H_{\theta_2})+\sum_{k\ge 1}\mca{D}[L_{k,\theta_1},L_{k,\theta_2}]\varrho_{t,\vb*{\theta}}\notag\\
	&\coloneqq\mca{L}_{t,\vb*{\theta}}(\varrho_{t,\vb*{\theta}}),
\end{align}
with the initial condition $\varrho_{0,\vb*{\theta}}=\varrho_0$.
Here, $\mca{D}[A,B]\circ\coloneqq A\circ B -(A^\dagger A\circ + \circ B^\dagger B)/2$. 
Consequently, the quantum Fisher information can be analytically calculated as 
\begin{equation}\label{eq:qFI.gen.form}
	\mca{I}_{\theta}=4\eval{\partial_{\theta_1\theta_2}^2\qty(\ln\qty|\tr\varrho_{\tau,\vb*{\theta}}|)}_{\theta_1=\theta_2=\theta},
\end{equation}
where $\varrho_{\tau,\vb*{\theta}}=\vec{\mca{T}}e^{\int_0^\tau\dd{t}\mca{L}_{t,\vb*{\theta}}}(\varrho_0)$ and $\vec{\mca{T}}$ denotes the time-ordering operator.

\subsection{Evaluation under perturbation}\label{app:proof.qtkur}
Here we evaluate $\mca{I}_0$ and $\partial_\theta\ev{\mca{J}_1}_\theta|_{\theta=0}$, and obtain the quantum TKUR for interacting quantum systems.
Thus far, the auxiliary dynamics is not specified yet.
Following the approach in Ref.~\cite{Vu.2025.PRXQ}, we consider the auxiliary dynamics in which the Hamiltonian remains unchanged and the jump operators are perturbed as
\begin{align}
	H_\theta &= H, \quad L_{\alpha_i k_i, \theta} = \sqrt{1 + \ell_{\alpha_i k_i}(t)\theta}L_{\alpha_i k_i},
\end{align}
where the coefficients $\{ \ell_{\alpha_i k_i}(t) \}$ are given by
\begin{equation}
	\ell_{\alpha_1 k_1}(t) =
		\dfrac{ \tr(L_{\alpha_1 k_1} \varrho_t L_{\alpha_1 k_1}^\dagger) - \tr(L_{\alpha_1 k_1^*} \varrho_t L_{\alpha_1 k_1^*}^\dagger) }{ \tr(L_{\alpha_1 k_1} \varrho_t L_{\alpha_1 k_1}^\dagger) + \tr(L_{\alpha_1 k_1^*} \varrho_t L_{\alpha_1 k_1^*}^\dagger) },
\end{equation}
and $\ell_{\alpha_ik_i}(t)=0$ for all $i>1$.
In other words, only the jump operators acting on subsystem $X_1$ are modified by parameter $\theta$ and the others remain unchanged.
Note that $\ell_{\alpha_1k_1}(t)$ is generally time-dependent as $\varrho_t$ may change over time.
When $\theta=0$, the auxiliary dynamics reduces exactly to the original one.
We evaluate the quantum Fisher information $\mca{I}_0$ for this auxiliary GKSL dynamics according to Eq.~\eqref{eq:qFI.gen.form} as
\begin{equation}
	\mca{I}_0=4\eval{\partial_{\theta_1\theta_2}^2\qty(\ln|\tr\varrho_{\tau,\vb*{\theta}}|)}_{\theta_1=\theta_2=0},
\end{equation}
where $\varrho_{\tau,\vb*{\theta}}=\vec{\mca{T}}e^{\int_0^\tau \dd{t} \mca{L}_{t,\vb*{\theta}}}(\varrho_0)$ is the operator evolved according to the following modified superoperator:
\begin{widetext}
\begin{align}
  \mca{L}_{t,\vb*{\theta}}(\circ)&=-i[H,\circ]+\sum_{i=1}^M\sum_{\alpha_i,k_i}\sqrt{[1+\ell_{\alpha_ik_i}(t)\theta_1][1+\ell_{\alpha_ik_i}(t)\theta_2]}L_{\alpha_ik_i}\circ L_{\alpha_ik_i}^\dagger \notag\\
  &-\frac{1}{2}\sum_{i=1}^M\sum_{\alpha_i,k_i}[1+\ell_{\alpha_i k_i}(t)\theta_1]L_{\alpha_i k_i}^\dagger L_{\alpha_i k_i}\circ -\frac{1}{2}\sum_{i=1}^M\sum_{\alpha_i, k_i}[1+\ell_{\alpha_i k_i}(t)\theta_2]\circ  L_{\alpha_i k_i}^\dagger L_{\alpha_i k_i},\label{eq:mod.op}
\end{align}
and $\varrho_0$ is the initial state of the system.
To calculate $\mca{I}_0$, it is convenient to use the vectorization representation, which vectorizes operators $A=\sum_{m,n}a_{mn}\dyad{m}{n}$ as $\kvec{A}\coloneqq\sum_{m,n}a_{mn}\ket{m}\otimes \ket{n}$.
By definition, it follows that $\kvec{AB}=(A\otimes \mbb{1})\kvec{B}$ and $\kvec{BA}=(\mbb{1} \otimes A^\top )\kvec{B}$.
Using this representation, the superoperator in Eq.~\eqref{eq:mod.op} can be vectorized as
\begin{align}
  \widehat{\mca{L}}_{t,\vb*{\theta}}&\coloneqq -i(H\otimes \mbb{1}_S-\mbb{1}_S\otimes H^\top)
  +\sum_{i=1}^M\sum_{\alpha_i,k_i}\sqrt{[1+\ell_{\alpha_ik_i}(t)\theta_1][1+\ell_{\alpha_ik_i}(t)\theta_2]}L_{\alpha_i k_i}\otimes L_{\alpha_i k_i}^*\notag
  \\
  &-\frac{1}{2}\sum_{i=1}^M\sum_{\alpha_i,k_i}[1+\ell_{\alpha_ik_i}(t)\theta_1]L_{\alpha_i k_i}^\dagger L_{\alpha_i k_i}\otimes \mbb{1}
  -\frac{1}{2}\sum_{i=1}^M\sum_{\alpha_i, k_i}[1+\ell_{\alpha_ik_i}(t)\theta_2]\mbb{1} \otimes (L_{\alpha_i k_i}^\dagger L_{\alpha_i k_i})^\top .
\end{align}
Here $\top$ and $*$ denote the matrix transpose and complex conjugate, respectively.
Using this representation, we have $\tr \varrho_{\tau,\vb*{\theta}}=\bvec{\mbb{1}_S}\vec{\mca{T}} e^{\int_0^\tau\dd{t}\widehat{\mca{L}}_{t,\vb*{\theta}}}\kvec{\varrho_0}$.
Note that the equality
\begin{align}
\partial_\theta e^{\int_0^\tau \dd{t} \widehat{\mca{L}}_{t,\theta}} =\vec{\mca{T}}\int_0^\tau \dd{t} e^{\int_t^\tau \dd{s} \widehat{\mca{L}}_{s,\theta}}\partial_{\theta} \widehat{\mca{L}}_{t,\theta}e^{\int_0^t \dd{s} \widehat{\mca{L}}_{s,\theta}}
\end{align}
holds for operators $\widehat{\mca{L}}_{t,\theta}$ dependent on parameter $\theta$.
Using this identity, the quantum Fisher information $\mca{I}_0$ can be calculated as follows:
\begin{align}
	\mca{I}_0&=4\eval{\qty[\partial_{\theta_1\theta_2}^2\bvec{\mbb{1}_S}\vec{\mca{T}} e^{\int_0^\tau \dd{t} \widehat{\mca{L}}_{t,\vb*{\theta}}}\kvec{\varrho_0} -\partial_{\theta_1}\bvec{\mbb{1}_S}\vec{\mca{T}}e^{\int_0^\tau \dd{t} \widehat{\mca{L}}_{t,\vb*{\theta}}}\kvec{\varrho_0} \partial_{\theta_2}\bvec{\mbb{1}_S}\vec{\mca{T}}e^{\int_0^\tau \dd{t} \widehat{\mca{L}}_{t,\vb*{\theta}}}\kvec{\varrho_0}]}_{\vb*{\theta}=\vb*{0}}\notag\\
	&=4\eval{\partial_{\theta_2}\bvec{\mbb{1}_S}\vec{\mca{T}}\int_0^\tau \dd{t} e^{\int_t^\tau \dd{s} \widehat{\mca{L}}_{s,\vb*{\theta}}}\partial_{\theta_1} \widehat{\mca{L}}_{t,\vb*{\theta}}e^{\int_0^t \dd{s} \widehat{\mca{L}}_{s,\vb*{\theta}}}\kvec{\varrho_0}}_{\vb*{\theta}=\vb*{0}}\notag\\
  &-4\eval{\bvec{\mbb{1}_S}\vec{\mca{T}}\int_0^\tau \dd{t} e^{\int_t^\tau \dd{s} \widehat{\mca{L}}_{s,\vb*{\theta}}}\partial_{\theta_1} \widehat{\mca{L}}_{t,\vb*{\theta}}e^{\int_0^t \dd{s} \widehat{\mca{L}}_{s,\vb*{\theta}}}\kvec{\varrho_0}\bvec{\mbb{1}_S}\vec{\mca{T}}\int_0^\tau \dd{t} e^{\int_t^\tau \dd{s} \widehat{\mca{L}}_{s,\vb*{\theta}}}\partial_{\theta_2} \widehat{\mca{L}}_{t,\vb*{\theta}}e^{\int_0^t \dd{s} \widehat{\mca{L}}_{s,\vb*{\theta}}}\kvec{\varrho_0}}_{\vb*{\theta}=\vb*{0}}\notag\\
  &=-4\eval{\bvec{\mbb{1}_S}\vec{\mca{T}}\int_0^\tau \dd{t} e^{\int_t^\tau \dd{s} \widehat{\mca{L}}_{s,\vb*{\theta}}}\partial_{\theta_1}\widehat{\mca{L}}_{t,\vb*{\theta}}e^{\int_0^t \dd{s} \widehat{\mca{L}}_{s,\vb*{\theta}}}\kvec{\varrho_0}\bvec{\mbb{1}_S}\hat{\mca{T}}\int_0^\tau \dd{t} e^{\int_t^\tau \dd{s} \widehat{\mca{L}}_{s,\vb*{\theta}}}\partial_{\theta_2} \widehat{\mca{L}}_{t,\vb*{\theta}}e^{\int_0^t \dd{s} \widehat{\mca{L}}_{s,\vb*{\theta}}}\kvec{\varrho_0}}_{\vb*{\theta}=\vb*{0}}\notag\\
  &+4\eval{\bvec{\mbb{1}_S}\hat{\mca{T}}\int_0^\tau\dd{t}\int_t^{\tau}\dd{t'} e^{\int_{t'}^\tau \dd{s} \widehat{\mca{L}}_{s,\vb*{\theta}}}\partial_{\theta_2} \widehat{\mca{L}}_{t',\vb*{\theta}}e^{\int_t^{t'} \dd{s} \widehat{\mca{L}}_{s,\vb*{\theta}}}\partial_{\theta_1} \widehat{\mca{L}}_{t,\vb*{\theta}}e^{\int_0^t \dd{s} \widehat{\mca{L}}_{s,\vb*{\theta}}}\kvec{\varrho_0}}_{\vb*{\theta}=\vb*{0}}\notag\\
  &+4\eval{\bvec{\mbb{1}_S}\vec{\mca{T}}\int_0^\tau\dd{t}\int_0^{t}\dd{t'} e^{\int_{t}^\tau \dd{s} \widehat{\mca{L}}_{s,\vb*{\theta}}}\partial_{\theta_1} \widehat{\mca{L}}_{t,\vb*{\theta}}e^{\int_{t'}^{t} \dd{s} \widehat{\mca{L}}_{s,\vb*{\theta}}}\partial_{\theta_2} \widehat{\mca{L}}_{t',\vb*{\theta}}e^{\int_0^{t'} \dd{s} \widehat{\mca{L}}_{s,\vb*{\theta}}}\kvec{\varrho_0}}_{\vb*{\theta}=\vb*{0}}\notag\\
  &+4\eval{\bvec{\mbb{1}_S}\vec{\mca{T}}\int_0^\tau \dd{t} e^{\int_t^\tau \dd{s} \widehat{\mca{L}}_{s,\vb*{\theta}}}\partial_{\theta_1\theta_2}^2 \widehat{\mca{L}}_{t,\vb*{\theta}}e^{\int_0^t \dd{s} \widehat{\mca{L}}_{s,\vb*{\theta}}}\kvec{\varrho_0}}_{\vb*{\theta}=\vb*{0}}\notag\\
  &=4\Big[-\bvec{\mbb{1}_S}\int_0^\tau \dd{t}\widehat{\mca{F}}_{1,t}\kvec{\varrho_t} \bvec{\mbb{1}_S}\int_0^\tau \dd{t}\widehat{\mca{F}}_{2,t}\kvec{\varrho_t}\notag\\
  &+\eval{\bvec{\mbb{1}_S}\vec{\mca{T}}\int_0^\tau\dd{t}\int_t^{\tau}\dd{t'} \widehat{\mca{F}}_{2,t'}\widehat{\mca{L}}_{t',\vb*{\theta}}e^{\int_t^{t'} \dd{s} \widehat{\mca{L}}_{s,\vb*{\theta}}}\widehat{\mca{F}}_{1,t}\kvec{\varrho_t}}_{\vb*{\theta}=\vb*{0}}\notag\\
  &+\eval{\bvec{\mbb{1}_S}\vec{\mca{T}}\int_0^\tau\dd{t}\int_0^{t}\dd{t'} \widehat{\mca{F}}_{1,t}e^{\int_{t'}^{t} \dd{s} \widehat{\mca{L}}_{s,\vb*{\theta}}}\widehat{\mca{F}}_{2,t'}\kvec{\varrho_{t'}}}_{\vb*{\theta}=\vb*{0}}\notag\\
  &+\eval{\bvec{\mbb{1}_S}\int_0^\tau \dd{t} \partial_{\theta_1\theta_2}^2 \widehat{\mca{L}}_{t,\vb*{\theta}}\kvec{\varrho_t}}_{\vb*{\theta}=\vb*{0}}
  \Big],
\end{align}
where we use the fact $\bvec{\mbb{1}_S}e^{\widehat{\mca{L}}t}=\bvec{\mbb{1}_S}$ and the operators $\widehat{\mca{F}}_{1,t}$ and $\widehat{\mca{F}}_{2,t}$ are defined as
\begin{align}
  \widehat{\mca{F}}_{1,t}&\coloneqq \eval{\partial_{\theta_1}\widehat{\mca{L}}_{t,\vb*{\theta}}}_{\vb*{\theta}
  =\vb*{0}}=\frac{1}{2}\sum_{i=1}^M\sum_{\alpha_i, k_i}\ell_{\alpha_i k_i}(t)\qty[ L_{\alpha_i k_i}\otimes L_{\alpha_ik_i}^* - (L_{\alpha_ik_i}^\dagger L_{\alpha_ik_i})\otimes\mbb{1}_S],\\
	\widehat{\mca{F}}_{2,t}&\coloneqq \eval{\partial_{\theta_2}\widehat{\mca{L}}_{t,\vb*{\theta}}}_{\vb*{\theta}=\vb*{0}}=\frac{1}{2}\sum_{i=1}^M\sum_{\alpha_i,k_i}\ell_{\alpha_ik_i}(t)\qty[ L_{\alpha_ik_i}\otimes L_{\alpha_ik_i}^* - \mbb{1}_S\otimes(L_{\alpha_ik_i}^\dagger L_{\alpha_ik_i} )^\top].
\end{align}
Since $\bvec{\mbb{1}_S}\widehat{\mca{F}}_{1,t}\kvec{A}=\bvec{\mbb{1}_S}\widehat{\mca{F}}_{2,t}\kvec{A}=0$ holds for any operator $A$, we obtain
\begin{align}
  \mca{I}_0&=4\eval{\bvec{\mbb{1}_S}\int_0^\tau \dd{t} \partial_{\theta_1\theta_2}^2 \widehat{\mca{L}}_{t,\vb*{\theta}}\kvec{\varrho_t}}_{\vb*{\theta}=\vb*{0}}\notag\\
  &=\eval{\bvec{\mbb{1}_S}\int_0^\tau \dd{t} \sum_{i=1}^M\sum_{\alpha_i,k_i}\ell_{\alpha_ik_i}^2(t)L_{\alpha_ik_i}\otimes L_{\alpha_ik_i}^*\kvec{\varrho_t}}_{\vb*{\theta}=\vb*{0}}\notag\\
  &=\int_0^\tau \dd{t} \sum_{\alpha_1,k_1}\ell_{\alpha_1k_1}^2 (t)\tr(L_{\alpha_1k_1}\varrho_t L_{\alpha_1k_1}^\dagger ).\label{eq:app.Fisher.eval}
\end{align}
We can show that the partial entropy production rate and partial dynamical activity associated with subsystem $X_1$ can be expressed as $\dot S_1^{\rm tot}=(1/2)\sum_{\alpha_1,k_1}\sum_{m,n}\sigma_{\alpha_1k_1}^{mn}$ and $\dot A_1=(1/2)\sum_{\alpha_1,k_1}\sum_{m,n}a_{\alpha_1k_1}^{mn}$, where
\begin{align}
	\sigma_{\alpha_1k_1}^{mn}&\coloneqq (w_{\alpha_1k_1}^{mn}p_n-w_{\alpha_1k_1^*}^{nm}p_m)\ln\frac{w_{\alpha_1k_1}^{mn}p_n}{w_{\alpha_1k_1^*}^{nm}p_m},\\
	a_{\alpha_1k_1}^{mn}&\coloneqq w_{\alpha_1k_1}^{mn}p_n+w_{\alpha_1k_1^*}^{nm}p_m.
\end{align}
Using these expressions, we can upper bound the quantum Fisher information by the partial entropy production and partial dynamical activity as follows:
\begin{align}
  \mca{I}_0&=\frac{1}{2}\int_0^\tau \dd{t}\sum_{\alpha_1, k_1}\frac{\qty[\tr(L_{\alpha_1 k_1}\varrho_t L_{\alpha_1k_1}^\dagger)-\tr(L_{\alpha_1k_1^*}\varrho_t L_{\alpha_1k_1^*}^\dagger)]^2}{\tr(L_{\alpha_1k_1}\varrho_t L_{\alpha_1k_1}^\dagger)+\tr(L_{\alpha_1k_1^*}\varrho_t L_{\alpha_1k_1^*}^\dagger)}\notag\\
  &=\frac{1}{2}\int_0^\tau \dd{t} \sum_{\alpha_1 k_1}\frac{\qty[\sum_{m,n}(w^{mn}_{\alpha_1k_1}p_n-w^{nm}_{\alpha_1k_1^*}p_m)]^2}{\sum_{m,n}(w^{mn}_{\alpha_1k_1}p_n+w^{nm}_{\alpha_1k_1^*}p_m)}\notag\\
  &\le \frac{1}{2}\int_0^\tau \dd{t}\sum_{\alpha_1,k_1}\sum_{m,n}\frac{\qty[(w^{mn}_{\alpha_1k_1}p_n-w^{nm}_{\alpha_1k_1^*}p_m)]^2}{w^{mn}_{\alpha_1k_1}p_n+w^{nm}_{\alpha_1k_1^*}p_m}\notag\\
  &= \int_0^\tau \dd{t} \sum_{\alpha_1,k_1,m,n}\frac{(\sigma^{mn}_{\alpha_1k_1})^2}{8a^{mn}_{\alpha_1k_1}}f\qty(\frac{{\sigma}^{mn}_{\alpha_1k_1}}{2a^{mn}_{\alpha_1k_1}})^{-2}\notag\\
  &\le \int_0^\tau \dd{t}\frac{(\dot{S}_1^{\rm tot})^2}{4\dot{A}_1}f\qty(\frac{\dot{S}_1^{\rm tot}}{2\dot{A}_1})^{-2}\notag\\
  &\le\frac{(\Delta S_1^{\rm tot})^2}{4 A_1} f\qty( \frac{\Delta S_1^{\rm tot}}{2 A_1} )^{-2}.\label{eq:app.qFI.ub}
\end{align}
Here we use the fact that $(x^2/y)f(x/y)^{-2}$ is a concave function and apply the Cauchy-Schwarz inequality and Jensen’s inequality.

Next, we calculate the term $\partial_\theta \ev{\mca{J}_1}_\theta$.
For $\theta \ll 1$, the density operator $\varrho_{t,\theta}$ in the auxiliary dynamics \eqref{eq:gen.mas} can be expanded in terms of $\theta$ as $\varrho_{t,\theta}=\varrho_t+\theta \varphi_t+O(\theta^2)$, where $\varrho_t=\varrho_{t,0}$.
Substituting this into Eq.~\eqref{eq:gen.mas} yields
\begin{align}
  \dot{\varrho}_t + \theta \dot{\varphi}_t &= -i \left[H, \varrho_t + \theta \varphi_t \right] + \sum_{i=1}^M \sum_{\alpha_i, k_i} [1 + \ell_{\alpha_ik_i}(t) \theta] \mca{D}[L_{\alpha_ik_i}] (\varrho_t + \theta \varphi_t) + \mathcal{O}(\theta^2).
\end{align}
By collecting the first-order terms, we obtain the differential equation that describes the time evolution of the operator $\varphi_t$,
\begin{align}
	\dot{\varphi}_t=-i[H,\varphi_t]+\sum_{i=1}^M \sum_{\alpha_i,k_i}\mca{D}[L_{\alpha_ik_i}]\varphi_t+\sum_{\alpha_1, k_1}\ell_{\alpha_1k_1}(t)\mca{D}[L_{\alpha_1k_1}]\varrho_t,
\end{align}
where the initial condition is given by $\varphi_0=\mbb{0}$.
It is evident that the operator $\varphi_t$ is always traceless.
Noting that $c_{\alpha_1k_1}=-c_{\alpha_1k_1^*}$, $c_{\alpha_ik_i}=0$ for $i>1$, and $c_{\alpha_1k_1}\ell_{\alpha_1k_1}=c_{\alpha_1k_1^*}\ell_{\alpha_1k_1^*}$, the partial derivative of the current average in the auxiliary dynamics with respect to $\theta$ can be calculated as
\begin{align}
	\eval{\partial_{\theta}\ev{\mca{J}_1}_\theta}_{\theta=0}&=\eval{\partial_{\theta}\qty[\int_0^\tau\dd{t}\sum_{\alpha_1,k_1}c_{\alpha_1k_1}[1+\ell_{\alpha_1k_1}(t)\theta]\tr{L_{\alpha_1k_1}(\varrho_t+\theta\varphi_t)L_{\alpha_1k_1}^\dagger}+O(\theta^2)]}_{\theta=0}\notag\\
	&=\int_0^\tau\dd{t}\sum_{\alpha_1,k_1}c_{\alpha_1k_1}\ell_{\alpha_1k_1}(t)\tr(L_{\alpha_1k_1}\varrho_t L_{\alpha_1k_1}^\dagger)+\int_0^\tau\dd{t}\sum_{\alpha_1,k_1}c_{\alpha_1k_1}\tr(L_{\alpha_1k_1}\varphi_t L_{\alpha_1k_1}^\dagger)\notag\\
	&=\frac{1}{2}\int_0^\tau\dd{t}\sum_{\alpha_1,k_1}c_{\alpha_1k_1}\qty[\tr(L_{\alpha_1k_1}\varrho_t L_{\alpha_1k_1}^\dagger)-\tr(L_{\alpha_1k_1^*}\varrho_t L_{\alpha_1k_1^*}^\dagger)]+\int_0^\tau\dd{t}\sum_{\alpha_1,k_1}c_{\alpha_1k_1}\tr(L_{\alpha_1k_1}\varphi_t L_{\alpha_1k_1}^\dagger)\notag\\
	&=\int_0^\tau\dd{t}\sum_{\alpha_1,k_1}c_{\alpha_1k_1}\tr(L_{\alpha_1k_1}\varrho_t L_{\alpha_1k_1}^\dagger)+\int_0^\tau\dd{t}\sum_{\alpha_1,k_1}c_{\alpha_1k_1}\tr(L_{\alpha_1k_1}\varphi_t L_{\alpha_1k_1}^\dagger)\notag\\
	&=\ev{\mca{J}_1}+\ev{\mca{J}_1}_\varphi,\label{eq:app.par.avg.TUR}
\end{align}
where we define $\ev{\mca{J}_1}_\varphi\coloneqq\int_0^\tau\dd{t}\sum_{\alpha_1,k_1}c_{\alpha_1k_1}\tr(L_{\alpha_1k_1}\varphi_t L_{\alpha_1k_1}^\dagger)$.
Defining the correction term:
\begin{equation}\label{eq:app.cur.avg.qcor}
	\delta_{\mca{J}_1}\coloneqq\frac{\ev{\mca{J}_1}_\varphi}{\ev{\mca{J}_1}},
\end{equation}
we readily obtain the quantum TKUR \eqref{eq:main.result.1} from Eqs.~\eqref{eq:app.Cramer-Rao1}, \eqref{eq:app.qFI.ub}, and \eqref{eq:app.par.avg.TUR},
\begin{equation}
	\frac{\Var[\mca{J}_1]}{\ev{\mca{J}_1}^2}\ge (1+\delta_{\mca{J}_1})^2\frac{4A_1}{(\Delta S_1^{\rm tot})^2}f\qty(\frac{\Delta S_1^{\rm tot}}{2A_1})^{2}.
\end{equation}

\subsection{Similar result for quantum diffusion unraveling}\label{app:qTKUR.qdu}
Here, we demonstrate that an analogous quantum TKUR can be derived for quantum diffusion unraveling using the same approach.

To this end, we first briefly outline the method of quantum diffusion unraveling (see Ref.~\cite{Landi.2024.PRXQ} for further details).
Unlike quantum jump unraveling, where the quantum state evolves through discontinuous jumps, quantum diffusion unraveling describes a continuous stochastic evolution due to a diffusion-like noise process.
In the short-time limit, the GKSL equation \eqref{eq:gen.mas} can also be expressed in the form of the Kraus representation as
\begin{equation}\label{eq:Kraus.rep}
	\varrho_{t+\Delta t,\theta}=\sum_{k\ge 0}M_{k,\theta}\varrho_{t,\theta}M_{k,\theta}^\dagger,
\end{equation}
where the Kraus operators are given by
\begin{equation}
    M_{0,\theta}=\mbb{1}_S-i\qty[H_\theta-\frac{i}{2}\sum_{k\ge 1}\qty{(\alpha_k^* L_{k,\theta} -\alpha_k L_{k,\theta}^\dagger)+(L_{k,\theta}+\alpha_k)^\dagger (L_{k,\theta}+\alpha_k)}]\Delta t,~M_{k,\theta}=(L_{k,\theta}+ \alpha_k)\sqrt{\Delta t}.
\end{equation}
Here, $\{\alpha_k\}_{k\ge 1}$, referred to as reference currents, are arbitrary constants.
Writing $\alpha_k=|\alpha_k|e^{i\phi_k}$, the rate of the average current can be expressed as
\begin{align}
  &\sum_{k\ge 1} c_k \tr[(L_{k,\theta}+\alpha_k)^\dagger (L_{k,\theta}+\alpha_k)\varrho_{t,\theta}] = \sum_{k\ge 1} c_k \qty(|\alpha_k|^2+|\alpha_k|\tr{(e^{-i\phi_k} L_{k,\theta} +e^{i\phi_k}L_{k,\theta}^\dagger)\varrho_{t,\theta}}+\tr{L_{k,\theta}^\dagger L_{k,\theta}\varrho_{t,\theta}}).\label{eq:diff.cur.tmp1}
\end{align}
We are interested in the regime where $\{|\alpha_k|\}_{k\ge 1}$ are sufficiently large. In this limit, the current is predominantly determined by the second term in Eq.~\eqref{eq:diff.cur.tmp1}, while the contribution from the last term becomes negligible. 
By subtracting the constant term and ignoring the relatively small term, we consider the normalized current: 
\begin{align}
  \ev{\dot{\mca{J}}}=\sum_{k\ge 1} c_k \tr[(e^{-i\phi_k} L_{k,\theta}+e^{i\phi_k}L_{k,\theta}^\dagger)\varrho_{t,\theta}],
\end{align}
which can be interpreted as the average of the following stochastic instantaneous current:
\begin{align}
  \dot{\mca{J}}=\sum_{k\ge 1} \frac{c_k}{|\alpha_k|} \qty(\dv{N_{k,\theta,t}}{t}-|\alpha_k|^2).
\end{align}
Here, $N_{k,\theta,t}$ denotes the number of occurrences of the quantum jump associated with the operator $L_{k,\theta}+\alpha_k$ within the time interval $[0,t]$.
At the trajectory level, the corresponding time-integrated current associated with each stochastic trajectory $\Gamma$ can be defined as
\begin{align}
  \mca{J}(\Gamma)\coloneqq \sum_{i=1}^N \frac{c_{k_i}}{|\alpha_{k_i}|}-\tau \sum_{k\ge 1} c_k |\alpha_k|.
\end{align}
Since the proof in Appendix \ref{app:gen.CR.ine} applies to arbitrary trajectory observables $\mca{J}(\Gamma)$, the generalized quantum Cram{\'e}r-Rao inequality \eqref{eq:gen.CR.ine} also holds in this case.
The quantum Fisher information can be calculated in the same way as for quantum jump unraveling.
This can be readily verified by noting that substituting the Kraus operators $\{M_{k,\theta}\}$ into Eq.~\eqref{eq:two.side.op.evol} reproduces Eq.~\eqref{eq:two.side.master.eq} exactly.

For simplicity, we consider real reference currents (i.e., $\phi_k=0$), while generalization to complex cases is straightforward.
In the $|\alpha_k|\rightarrow \infty$ limit, the GKSL dynamics \eqref{eq:gen.mas} can be unraveled into a stochastic evolution of the conditioned density matrix $\psi_{t,\theta}$, described by the following It{\^o} stochastic differential equation:
\begin{equation}
	d\psi_{t,\theta}=\qty(-i[H_{\theta},\psi_{t,\theta}]+\sum_{k\ge 1}\mca{D}[L_{k,\theta}]\psi_{t,\theta})dt
  +\sum_{k\ge 1}\qty[L_{k,\theta}\psi_{t,\theta} + \psi_{t,\theta}L_{k,\theta}^\dagger - \tr(L_{k,\theta}\psi_{t,\theta} + \psi_{t,\theta}L_{k,\theta}^\dagger)\psi_{t,\theta}]dW_{k,t}.
\end{equation}
Here, $H_{{\rm eff},\theta}\coloneqq H_{\theta}-(i/2)\sum_{k\ge 1}L_{k,\theta}^\dagger L_{k,\theta}$, and $\{dW_{k,t}\}$ are the independent Wiener increments satisfying $\mbb{E}[dW_{k,t}]=0$ and $\mbb{E}[dW_{k,t}dW_{k',t}]=\delta_{kk'}dt$.
Such a stochastic differential equation is called the quantum state diffusion equation \cite{Breuer.2002}.
Since the rate $dN_{k,\theta,t}/dt$ behaves as \cite{Landi.2024.PRXQ}
\begin{align}
  \dv{N_{k,\theta,t}}{t}\simeq |\alpha_k|^2 +|\alpha_k|\qty(\ev{L_{k,\theta} +L_{k,\theta}^\dagger}_{t,\theta}+\dv{W_{k,t}}{t}),
\end{align}
the stochastic instantaneous current can be written as
\begin{align}
  \dot{\mca{J}}&=\sum_{k\ge 1} c_k \qty(\ev{L_{k,\theta} +L_{k,\theta}^\dagger}_{t,\theta}+\dv{W_{k,t}}{t}).
\end{align}

We now return to the problem of deriving the quantum TKUR for subsystems within interacting quantum systems. At the trajectory level, the stochastic current $\mca{J}_1$ occurring within subsystem $X_1$ is given by
\begin{equation}
	\mca{J}_1=\int_0^\tau\sum_{\alpha_1,k_1}c_{\alpha_1k_1}\qty(\ev{L_{\alpha_1k_1}+L_{\alpha_1k_1}^\dagger}_t\dd{t}+dW_{\alpha_1k_1,t}).
\end{equation}
Its ensemble average can be calculated as
\begin{equation}
	\ev{\mca{J}_1}=\int_0^\tau\dd{t}\sum_{\alpha_1, k_1}c_{\alpha_1 k_1}\tr[(L_{\alpha_1 k_1}+L_{\alpha_1 k_1}^\dagger)\varrho_t].
\end{equation}
As explained above, the quantum Cram{\'e}r-Rao inequality \eqref{eq:gen.CR.ine} remains valid even for quantum diffusion unraveling. The key distinction from the case of quantum jump unraveling lies in the partial derivative of the current. In this case, it can be evaluated as follows: 
\begin{align}
\eval{\partial_{\theta}\ev{\mca{J}_1}_\theta}_{\theta=0}
	&=\eval{\partial_{\theta}\qty[\int_0^\tau\dd{t}\sum_{\alpha_1, k_1}c_{\alpha_1k_1}[1+\ell_{\alpha_1k_1}(t)\theta/2]\tr[(L_{\alpha_1k_1}+L_{\alpha_1k_1}^\dagger)(\varrho_t+\theta\varphi_t)]+O(\theta^2)]}_{\theta=0}\notag\\
	&=\frac{1}{2}\int_0^\tau\dd{t}\sum_{\alpha_1,k_1}c_{\alpha_1k_1}\ell_{\alpha_1k_1}(t)\tr[(L_{\alpha_1k_1}+L_{\alpha_1k_1}^\dagger)\varrho_t]+\int_0^\tau\dd{t}\sum_{\alpha_1,k_1}c_{\alpha_1k_1}\tr[(L_{\alpha_1k_1}+L_{\alpha_1k_1}^\dagger)\varphi_t]\notag\\
	&=\ev{\mca{J}_1}+\ev{\mca{J}_1}_*+\ev{\mca{J}_1}_\varphi,\label{eq:qdu.par.avg.TUR}
\end{align}
where $\ev{\mca{J}_1}_*\coloneqq \int_0^\tau\dd{t} \sum_{\alpha_1,k_1}c_{\alpha_1k_1}[\ell_{\alpha_1k_1}(t)/2-1]\tr[(L_{\alpha_1k_1}+L_{\alpha_1k_1}^\dagger)\varrho_t]$.
Defining the correction term $\delta_{\mca{J}_1}'\coloneqq(\ev{\mca{J}_1}_*+\ev{\mca{J}_1}_\varphi)/\ev{\mca{J}_1}$, we obtain the following quantum TKUR for quantum diffusion unraveling:
\begin{equation}
	\frac{\Var[\mca{J}_1]}{\ev{\mca{J}_1}^2}\ge (1+\delta_{\mca{J}_1}')^2\frac{4A_1}{(\Delta S_1^{\rm tot})^2}f\qty(\frac{\Delta S_1^{\rm tot}}{2A_1})^{2}.
\end{equation}
This relation has the same structure as Eq.~\eqref{eq:main.result.1} derived for quantum jump unraveling, differing only in the correction term $\delta_{\mca{J}_1}'$.

\section{Survival of the correction term in the fast-relaxation limit}\label{app:fast.relaxation}
We demonstrate that the correction term can remain finite even in the fast-relaxation limit.
For clarity, we consider a bipartite system composed of two subsystems $X$ and $Y$.
In the classical case, it was shown that when subsystem $Y$ relaxes much faster than $X$, the correction term $\delta_{\mca{J}_X}$ vanishes in the long-time limit $\tau\to\infty$ \cite{Tanogami.2023.PRR}.
However, we show that this does not generally hold for quantum systems.
For simplicity, we consider the steady-state regime, where both Hamiltonians and jump operators are time-independent.
We assume that the Lindbladian superoperator $\mca{L}$ is separated with respect to a parameter $\gamma \gg 1$ as $\mca{L}=\mca{L}_0+\gamma \mca{L}_1$, where
\begin{align}
	\mca{L}_0[\circ]&\coloneqq -i\comm{H}{\circ}+\sum_{\alpha_X,k_X}\mca{D}[L_{\alpha_Xk_X}]\circ,
  \\
  \mca{L}_1[\circ]&\coloneqq \sum_{\alpha_Y,k_Y}\mca{D}[L_{\alpha_Yk_Y}]\circ,
\end{align}
and $\{L_{\alpha_Yk_Y}\}$ are local jump operators that act only on subsystem $Y$ as $L_{\alpha_Yk_Y}=\mbb{1}_X\otimes L_{\alpha_Yk_Y}'$.
The condition $\gamma\gg 1$ ensures that subsystem $Y$ relaxes rapidly relative to $X$.
In the long-time limit, both operators $\varrho_t$ and $\varphi_t$ approach steady states $\varrho_{\rm ss}$ and $\varphi_{\rm ss}$, respectively, satisfying $\mca{L}[\varrho_{\mrm{ss}}]=\mbb{0}$ and $\mca{L}[\varphi_{\mrm{ss}}]=-\mca{D}_\ell [\varrho_{\mrm{ss}}]$, where $\mca{D}_\ell [\circ]\coloneqq \sum_{\alpha_X, k_X}\ell_{\alpha_X k_X}\mca{D}[L_{\alpha_X k_X}]\circ$.
Assuming that $\mca{L}[\circ]=\mbb{0}$ has a unique nontrivial solution, the analytical expression of $\varphi_{\mrm{ss}}$ can be calculated using the Moore-Penrose pseudo-inverse of $\mca{L}$ \cite{Vu.2025.PRXQ}.
In the vectorization representation, the counterpart of ${\mca{L}}$ can be expressed as $\widehat{\mca{L}}=\sum_{i>0}\chi_i\kvec{r_i}\bvec{l_i}$, where $0=\chi_0>\mrm{Re}(\chi_1)\geq\mrm{Re}(\chi_2)\geq\dots$ are the eigenvalues of $\widehat{\mca{L}}$, and $\kvec{r_i}$ and $\kvec{l_i}$ are its right and left eigenvectors, respectively.
Here, $\widehat{\mca{X}}$ denotes the vectorization representation of any superoperator $\mca{X}$.
The eigenvectors satisfy the relation $\kvec{\rho_{\mrm{ss}}}\bvec{\mbb{1}}+\sum_{i>0}\kvec{r_i}\bvec{l_i}=\mbb{1}$.
The Moore-Penrose pseudo-inverse of $\widehat{\mca{L}}$ is defined as $\widehat{\mca{L}}^+\coloneqq\sum_{i>0}\chi_i^{-1} \kvec{r_i}\bvec{l_i}$, which fulfills the relation $\widehat{\mca{L}}^+\widehat{\mca{L}}=\sum_{i>0}\kvec{r_i}\bvec{l_i}=\mbb{1}-\kvec{\rho_{\mrm{ss}}}\bvec{\mbb{1}}$.
Then, by noting that $\bkgvec{\mbb{1}}{\varphi_{\mrm{ss}}}=0$, $\varphi_{\rm ss}$ can be calculated from the relation $\widehat{\mca{L}}\kvec{\varphi_{\mrm{ss}}}=-\widehat{\mca{D}}_\ell \kvec{\varrho_{\mrm{ss}}}$ as $\kvec{\varphi_{\mrm{ss}}}=-\widehat{\mca{L}}^+ \widehat{\mca{D}}_\ell \kvec{\varrho_{\mrm{ss}}}$.
Using this formulation, the correction term in the long-time limit can be calculated as
\begin{equation}
    \delta_{\mca{J}_X}=\frac{\sum_{\alpha_X,k_X}c_{\alpha_Xk_X}\tr(L_{\alpha_Xk_X}\varphi_{\rm ss}L_{\alpha_Xk_X}^\dagger)}{\sum_{\alpha_X,k_X}c_{\alpha_Xk_X}\tr(L_{\alpha_Xk_X}\varrho_{\rm ss}L_{\alpha_Xk_X}^\dagger)}.
\end{equation}

Next, we expand both $\varrho_{\rm ss}$ and $\varphi_{\rm ss}$ in powers of $\gamma^{-1}$ as
\begin{align}
  \varrho_{\rm{ss}}=\varrho^{(0)}+\gamma^{-1}\varrho^{(1)}+\gamma^{-2}\varrho^{(2)}+\dots,
  \\
  \varphi_{\rm{ss}}=\varphi^{(0)}+\gamma^{-1}\varphi^{(1)}+\gamma^{-2}\varphi^{(2)}+\dots.
\end{align}
Similarly, coefficients $\{\ell_{\alpha_X k_X}\}$, defined using $\varrho_{\rm{ss}}$, can also be expanded in terms of $\gamma$ as $\ell_{\alpha_X k_X}=\ell_{\alpha_X k_X}^{(0)}+\gamma^{-1}\ell_{\alpha_X k_X}^{(1)}+\gamma^{-2}\ell_{\alpha_X k_X}^{(2)}+\dots$.
For convenience, we define $\mca{D}_\ell^{(i)}[\circ] \coloneqq \sum_{\alpha_X, k_X} \ell_{\alpha_X k_X}^{(i)}\mca{D}[L_{\alpha_X k_X}] \circ$.
By collecting terms of each order of $\gamma$ in the equation $\mca{L}[\varphi_{\mrm{ss}}]=-\mca{D}_\ell [\varrho_{\mrm{ss}}]$, we obtain a hierarchy of equations as follows:
\begin{align}
  \mca{L}_1[\varphi^{(0)}]&=\mbb{0},\label{app:eq.phi.1}
  \\
  \mca{L}_0[\varphi^{(0)}]+\mca{L}_1[\varphi^{(1)}]&=-\mca{D}_\ell^{(0)}[\varrho^{(0)}],\label{app:eq.phi.0}
  \\
  \mca{L}_0[\varphi^{(1)}]+\mca{L}_1[\varphi^{(2)}]&=-(\mca{D}_\ell^{(0)} [\varrho^{(1)}]+\mca{D}_\ell^{(1)}[\varrho^{(0)}]),
  \\
  \dots .
\end{align}
We show that $\varphi^{(0)} \neq \mbb{0}$ generally holds in quantum systems.
To this end, assume $\varphi^{(0)} = \mbb{0}$, which evidently satisfies the leading-order equation \eqref{app:eq.phi.1}.
Consequently, the next-order equation \eqref{app:eq.phi.0} gives $\mca{L}_1[\varphi^{(1)}] =-\mca{D}_\ell^{(0)}[\varrho^{(0)}]$.
Taking the partial trace of this equation over subsystem $Y$ yields $\mbb{0}=-\tr_Y(\mca{D}_\ell^{(0)}[\varrho^{(0)}])$, where the locality of the superoperator $\mca{L}_1$ is exploited (i.e., $\tr_Y(\mca{L}_1[\circ])=\mbb{0}$).
However, the right-hand side of this equation is generally nonzero, leading to a contradiction.
This implies that $\varphi^{(0)}\neq\mbb{0}$, and consequently, the correction term $\delta_{\mca{J}_X}$ may remain finite even in the fast-relaxation limit $\gamma\to \infty$.
In contrast, in the classical limit, $\tr_Y(\mca{D}_\ell^{(0)}[\varrho^{(0)}])=\mbb{0}$ holds, and the correction term vanishes.

\section{Recovery of known results from main relations \eqref{eq:main.result.1} and \eqref{eq:main.result.2} in several limiting cases}\label{app:special}
In what follows, we demonstrate that our main results, Eqs.~\eqref{eq:main.result.1} and \eqref{eq:main.result.2}, recover several previously established results in relevant limiting cases.

\subsection{The classical bipartite limit}
First, we show that our result \eqref{eq:main.result.2} reduces exactly to the TUR derived in Ref.~\cite{Tanogami.2023.PRR} for classical bipartite systems when taking the classical limit.
Consider a bipartite system composed of two subsystems, $X$ and $Y$, without any interaction Hamiltonian (i.e., $H_{\rm int} = \mbb{0}$). The subsystem Hamiltonians and jump operators are given by
\begin{align}
    H_X&=\sum_x \varepsilon_x^X \dyad{x},\\
    H_Y&=\sum_y \varepsilon_y^Y \dyad{y},\\
    L_{\alpha_Xk_X}&=\sqrt{\gamma_{xx'}^y}\dyad{x}{x'}\otimes\dyad{y}\eqqcolon L_{xx'}^y,\\
    L_{\alpha_Yk_Y}&=\sqrt{\gamma_{x}^{yy'}}\dyad{x}{x}\otimes\dyad{y}{y'}\eqqcolon L_{x}^{yy'}.
\end{align}
Here, $\{\varepsilon_x^X,\varepsilon_y^Y\}$ are energy levels, $\{\ket{x}\}$ and $\{\ket{y}\}$ are eigenstates of $H_X$ and $H_Y$, respectively, and $\{\gamma_{xx'}^y,\gamma_{x}^{yy'}\}$ are nonnegative transition rates.
In this case, the GKSL equation becomes
\begin{equation}
    \dot{\varrho}_t =-i[H_X+H_Y,\varrho_t]+\sum_y\sum_{x\neq x'}\mca{D}[L^y_{xx'}]\varrho_t+\sum_x\sum_{y\neq y'}\mca{D}[L^{yy'}_{x}]\varrho_t.
\end{equation}
Assuming the initial density matrix $\varrho_0$ is diagonal in the energy eigenstates, the time-evolved state $\varrho_t$ remains diagonal at all times: $\varrho_t=\sum_{x,y}p_t(x,y)\dyad{x,y}$, so that $[H_X+H_Y,\varrho_t]=\mbb{0}$.
Since $\mel{x'',y''}{\mca{D}[L^{y}_{xx'}]\varrho_t}{x'',y''}=\delta_{x'' x}\delta_{y''y}\gamma^y_{xx'}p_t(x',y)-\delta_{x'' x'}\delta_{y''y}\gamma^y_{xx'}p_t(x',y)$, the time evolution of the population probability is described by the master equation as
\begin{equation}\label{eq:bip.meq}
    \dot{p}_t(x,y) =\sum_{x'}\gamma^y_{xx'}p_t(x',y)+\sum_{y'}\gamma_x^{yy'}p_t(x,y'),
\end{equation}
where $\gamma^y_{xx}\coloneqq -\sum_{x'(\neq x)}\gamma^y_{x'x}$ and $\gamma^{yy}_{x}\coloneqq -\sum_{y'(\neq y)}\gamma^{y'y}_{x}$.
The obtained equation \eqref{eq:bip.meq} is exactly the master equation for classical bipartite systems.
Similarly, the operator $\varphi_t$ is also diagonal at all times, $\varphi_t=\sum_{x,y}q_t (x,y)\dyad{x,y}$, and $q_t(x,y)$ is governed by the equation:
\begin{align}
    \dot{q}_t (x,y)&=\sum_{x'}\gamma^y_{xx'}q_t(x',y)+\sum_{y'}\gamma^{yy'}_{x}q_t(x,y')+\sum_{x'(\neq x)}[\ell_{xx'}^y(t) \gamma^y_{xx'}p_t(x',y) - \ell_{x'x}^y(t) \gamma^y_{x'x}p_t(x,y)]\notag
    \\
    &=\sum_{x'}\gamma^y_{xx'}q_t(x',y)+\sum_{y'}\gamma^{yy'}_{x}q_t(x,y')+\sum_{x'}\gamma^y_{xx'}p_t(x',y),\label{eq:q.meq}
\end{align}
where
\begin{align}
    \ell_{xx'}^y(t)=\frac{\gamma_{xx'}^yp_t(x',y)-\gamma_{x'x}^yp_t(x,y)}{\gamma_{xx'}^yp_t(x',y)+\gamma_{x'x}^yp_t(x,y)}.
\end{align}
Both Eqs.~\eqref{eq:bip.meq} and \eqref{eq:q.meq} are the same as those in Ref.~\cite{Tanogami.2023.PRR}, and the correction term $\delta_{\mca{J}_1}$ likewise coincides.
Therefore, in the classical limit, our result \eqref{eq:main.result.2} correctly reproduces the TUR for classical bipartite systems.

\subsection{The noninteracting and local-jump limit}
Next, we show that our result reduces to the TKUR derived in Ref.~\cite{Vu.2025.PRXQ} for a single quantum system when there is no interaction between $X_1$ and the rest of the system, and all jump operators are local.
Specifically, we consider the case where $H_{\rm int}=\mbb{0}$ and the jump operators are of the form $L_{\alpha_1k_1}=L_{\alpha_1k_1}'\otimes\mbb{1}_{\setminus 1}$ and $L_{\alpha_ik_i}=\mbb{1}_1\otimes L_{\alpha_ik_i}'$ for $i>1$.
In this setting, the GKSL equation becomes
\begin{equation}
    \dot{\varrho}_t =-i[H_{X_1}+\dots+H_{X_M},\varrho_t]+\sum_{\alpha_1, k_1}\mca{D}[L_{\alpha_1 k_1}'\otimes \mbb{1}_{\setminus 1}]\varrho_t+\sum_{i=2}^M\sum_{\alpha_i, k_i}\mca{D}[\mbb{1}_1 \otimes L_{\alpha_i k_i}']\varrho_t.
\end{equation}
Taking the partial trace over $X_2,\dots,X_M$ yields the following equation for $\varrho_{X_1}(t)\coloneqq\tr_{\setminus 1}\varrho_t$:
\begin{equation}\label{eq:sub.meq}
    d_t\varrho_{X_1}(t) =-i[H_{X_1},\varrho_{X_1}(t)]+\sum_{\alpha_1, k_1}\mca{D}[L_{\alpha_1 k_1}']\varrho_{X_1}(t).
\end{equation}
Here, we use the fact that, for any operator $A$ of the Hilbert space $\mca{H}_X\otimes \mca{H}_Y$ and any operator $B$ of $\mca{H}_Y$, $\tr_Y(A \mbb{1}_X\otimes B)=\tr_Y(\mbb{1}_X\otimes B A)$ holds.
This means that, in the absence of interactions and nonlocal jumps, the dynamics of $X_1$ is governed solely by local terms, reducing to a single open quantum system.
Similarly, the dynamics of the operator $\varphi_t$ is given by 
\begin{equation}
    \dot{\varphi}_t =-i\qty[\sum_{i=1}^MH_{X_i},\varphi_t]+\sum_{\alpha_1, k_1}\mca{D}[L_{\alpha_1 k_1}'\otimes \mbb{1}_{\setminus 1}]\varphi_t+\sum_{i>1}\sum_{\alpha_i, k_i}\mca{D}[\mbb{1}_{1} \otimes L_{\alpha_i k_i}']\varphi_t
    +\sum_{\alpha_1, k_1}\ell_{\alpha_1 k_1}(t) \mca{D}[L_{\alpha_1 k_1}'\otimes \mbb{1}_{\setminus 1}]\varrho_t,
\end{equation}
and the time-evolution equation for $\varphi_{X_1}(t)\coloneqq\tr_{\setminus 1}\varphi_t$ can be immediately obtained as
\begin{equation}\label{eq:sub.phi.meq}
    \partial_t\varphi_{X_1}(t) =-i\qty[H_{X_1},\varphi_{X_1}(t)]+\sum_{\alpha_1, k_1}\mca{D}[L_{\alpha_1 k_1}']\varphi_{X_1}(t)
    +\sum_{\alpha_1, k_1}\ell_{\alpha_1 k_1}(t) \mca{D}[L_{\alpha_1 k_1}']\varrho_{X_1}(t).
\end{equation}
As can be seen, both Eqs.~\eqref{eq:sub.meq} and \eqref{eq:sub.phi.meq} are identical with those in Ref.~\cite{Vu.2025.PRXQ} for a single system.
Consequently, the correction term $\delta_{\mca{J}_1}$ also coincides, and our result \eqref{eq:main.result.1} reduces to the known TKUR for a single quantum system \cite{Vu.2025.PRXQ}.
\end{widetext}

\twocolumngrid

\end{document}